\documentclass[cmp]{svjour}
\usepackage{graphicx, amsmath, amsfonts}
\renewcommand{\vec}[1]{{\mathbf #1}}
\renewcommand{\Re}{{\mathfrak{Re}}}
\renewcommand{\Im}{{\mathfrak{Im}}}
\newcommand{\eref}[1]{(\ref{#1})}
\newcommand{\rmd}{{\mathrm d}}
\newcommand{\rme}{{\mathrm e}}
\newcommand{\rmi}{{\mathrm i}}
\newcommand{\Or}{{\mathrm O}}
\newcommand{\Lint}{\int_{\Lbar}^{\Lbar+\Delta L}}
\newcommand{\Lbar}{\bar{L}}
\newcommand{\erf}{\mathop{\rm erf}}
\newcommand{\erfc}{\mathop{\rm erfc}}

\newcommand{\meas}{\mathop{\rm meas}}
\newcommand{\I}{{1\!\!1}}
\newcommand{\vecxi}{{\boldsymbol \xi}}

\newcommand{\dimonstrazione}{\par\noindent{\sl Proof.}\phantom{X}}
\newcommand{\dimon}[1]{\par\noindent{\sl #1}\phantom{X}}
\newcommand{\finire}{\hfill$\square$\par\vspace{5mm}}
\begin{document}
\titlerunning{Eigenfunctions and spectral determinants of star graphs}
\title{Value distribution of the eigenfunctions and spectral determinants
of quantum star graphs}
\author{J. P. Keating, J. Marklof and B. Winn}
\institute{School of Mathematics, University of Bristol, Bristol. BS8 1TW, 
U.K.\\
\email{{j.p.keating@bristol.ac.uk}, {j.marklof@bristol.ac.uk},
{b.winn@bristol.ac.uk}}}
\date{Received: 29 October 2002}
\maketitle
\begin{abstract} We compute the value distributions of the eigenfunctions and
spectral determinant of the Schr\"odinger operator on families of star 
graphs. The values of the spectral determinant are shown to have a Cauchy 
distribution with respect both to averages over bond lengths in the limit as
the wavenumber tends to infinity and to averages over wavenumber when the
bond lengths are fixed and not rationally related. 
This is in contrast to the spectral determinants
of random matrices, for which the logarithm is known to satisfy a
Gaussian limit distribution. The value distribution of the eigenfunctions
also differs from the corresponding random matrix result. We argue that
the value distributions of the spectral determinant and of the eigenfunctions
should coincide with those of \v{S}eba-type billiards.
\end{abstract}
\section{Introduction}
The study of quantum graphs as model systems for quantum chaos was initiated
by Kottos and Smilansky \cite{kot:1}, \cite{kot:pot}, who observed that
the spectral statistics of fully-connected graphs are typical of those
associated with generic classically chaotic systems. The relative simplicity
of quantum graphs, together with the existence of an exact trace formula,
has lead to the suggestion that their study might provide 
insights into some of the fundamental problems of quantum chaos \cite{kot:2}. 
This has motivated many
works considering a variety of aspects of quantum graphs,
\cite{bar:lsd}, \cite{berk:3}, \cite{bolte:1}, \cite{desbois:1},
\cite{kurasov:1}, \cite{prot:1}, \cite{pascaud:1}, \cite{schanz:1},
\cite{schanz:2}, \cite{tanner:1}.

Studies \cite{berk:2}, \cite{berk:1} of a special class of graphs - 
the so-called ``hydra'' graphs or star graphs - have revealed spectral 
statistics 
that are not typically associated with quantum chaotic systems.
These have been dubbed ``intermediate statistics'' in recent works,
\cite{bogomolny:1}, \cite{bogomolny:3}, \cite{bogomolny:2} and have been 
observed in a number of systems. We are
motivated to investigate this further by studying the value distributions of
the eigenfunctions and the spectral determinant of
the Schr\"odinger operator on quantum star graphs.

\begin{figure}[h]
\begin{center}
\includegraphics[angle=0,width=3.0cm,height=2.5cm]{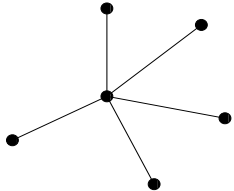}
\caption{A star graph with 5 bonds}
\label{fig:0}
\end{center}
\end{figure}

A {\em star graph} consists of a single central vertex together with $v$ 
outlying
vertices each of which is connected only to the central vertex by a bond
(figure \ref{fig:0}). Hence
there are $v$ bonds. We associate to each bond a length $L_j, j=1,\ldots,v$.
We will often refer to the vector of bond lengths $\vec{L}:=(L_1,\ldots,L_v)$.

The Schr\"odinger operator on a star graph takes the form of the Laplacian
$-\rmd^2/\rmd x^2$ acting on the space of functions defined on
the bonds of the graph that are twice-differentiable and satisfy the 
following matching conditions at the vertices:
\begin{eqnarray*}
\psi_j(0)=\psi_i(0)&=:&\Psi,\qquad j,i=1,\ldots,v\\
\sum_{j=1}^v \psi_j^{\prime}(0)&=&\frac{1}{\lambda}\Psi\\
\psi_j^{\prime}(L_j)&=&0,\qquad j=1,\ldots,v.
\end{eqnarray*}
Here $\psi_j$ is the component of the function defined on the $j^{\rm th}$ bond
of the graph, and $\psi_j:[0,L_j]\to{\mathbb R}$ with the convention that
$\psi_j(0)$ is the value of the function at the central vertex of the
star graph. $\lambda$ is a parameter that allows us to vary the boundary
conditions at the central vertex.

The Schr\"odinger operator so-defined is self-adjoint, so there exists a
discrete unbounded set of values $0\leq k_0 < k_1 \leq k_2 \leq \cdots \to
\infty$ such that
$k_n^2$ is an eigenvalue. It can be shown that $k=\pm k_n$ corresponds to an
eigenvalue if and only if it is a solution of $Z(k,\vec{L})=0$, where
\begin{equation}
Z(k,\vec{L}):=\sum_{j=1}^v \tan kL_j-\frac{1}{k\lambda}.
\label{spec:det}
\end{equation}
We refer to $Z(k,\vec{L})$ as the {\em spectral determinant}. Note that
$Z(k,\vec{L})$ has poles at $k=(2n+1)\pi/2L_j$ for each $n\in{\mathbb Z}$ and
$j=1,\ldots,v$. The zeros and poles of $Z(k,\vec{L})$ interlace.
The usual definition of the spectral determinant would require 
these poles to be factored out.

For simplicity we 
henceforth consider the case 
$1/\lambda=0$. We shall employ the notation
\begin{equation*}
  Z'(k,\vec{L})=\frac{\partial Z}{\partial k}(k,\vec{L})=\sum_{j=1}^v
L_j\sec^2 kL_j.
\end{equation*}

The eigenfunction corresponding to the $n^{\rm th}$ eigenvalue is found to be
\begin{equation*}
\psi_i^{(n)}(x)=A^{(n)}\frac{\cos k_n(x-L_i)}{\cos k_nL_i}.
\end{equation*}
The constant $A^{(n)}$ is determined by the normalisation 
\begin{equation}
\sum_{j=1}^v \int_0^{L_j} |\psi_j^{(n)}(x)|^2 \rmd x=1,
\label{norm:const}
\end{equation}
to be
\begin{equation*}
  A^{(n)}=\left(\frac{2}{\sum_{j=1}^v L_j\sec^2 k_nL_j}\right)^{\frac{1}{2}}.
\end{equation*}
The value distribution of the eigenfunctions is 
determined by these normalisation constants. For definiteness, we shall
focus here on the maximum amplitude
squared of the eigenfunctions on a single bond,
\begin{eqnarray*}
A_i(n,\vec{L};v)&:=&\sup_{x\in[0,L_i]}\{|\psi_{i}^{(n)}(x)|^2\}\\
&=&(A^{(n)}\sec k_nL_i)^2\\
&=&\frac{2\sec^2 k_nL_i}{\sum_{j=1}^v L_j\sec^2 k_nL_j}.
\end{eqnarray*}

We now state our main results.
\begin{theorem}
\label{thm:1}
For any fixed $\Lbar > 0$ 
\begin{equation*}
\lim_{k\to\infty} \frac{1}{(\Delta L)^v}
\meas\left\{ \vec{L}\in[\Lbar,\Lbar+\Delta L]^v:\frac{1}{v}Z(k,\vec{L})<y
\right\}=\frac{1}{\pi}
\int_{-\infty}^y \frac{1}{1+x^2} \rmd x,
\end{equation*}
provided 
that $k\Delta L\to\infty$ as $k\to\infty$.
\end{theorem}
We emphasise that
in theorem \ref{thm:1} we do not require that $\Delta L\to 0$. 
In some later results we shall make this stipulation.
\begin{theorem}
\label{thm:2}
Suppose that the components of $\vec{L}$ are fixed and linearly
independent over ${\mathbb Q}$. Then
\begin{equation*}
\lim_{K\to\infty} \frac{1}{K}\meas\left\{k\in[0,K]:\frac{1}{v}Z(k,\vec{L})<y
\right\}=\frac{1}{\pi}\int_{-\infty}^y\frac{1}{1+x^2}\rmd x.
\end{equation*}
\end{theorem}

Theorems \ref{thm:1} and \ref{thm:2} demonstrate the equivalence of taking a
$k$-average and a bond-length average at large $k$ for the distribution of
values taken by the function $Z(k,\vec{L})$. Such a correspondence was noted
in \cite{bar:lsd} for the spacing distribution of 
the eigenvalues of quantum graphs.

In \cite{snaith:1} it was shown that the value distributions of the real and 
imaginary parts of the logarithm of the characteristic polynomial of
a random unitary matrix drawn from the circular ensembles of random matrix 
theory tend independently to a Gaussian distribution in the limit as the
matrix size tends to infinity, subject to
appropriate normalisation. The distribution that appears in
theorems \ref{thm:1} and \ref{thm:2} is known as the Cauchy distribution.
It is related to the Gaussian distribution by the fact that both are examples
of a larger class of distributions known as stable distributions. Such 
distributions share the property that the sum of two random variables from
a stable distribution is distributed like a random variable from the
same distribution. 
Theorem \ref{thm:1} is a consequence of this fact.
We also note that the density in theorems \ref{thm:1} and
\ref{thm:2} is independent of $v$ when $Z(k,\vec{L})$ is normalised as 
indicated.

We next consider the distribution of values taken by $Z'(k)$
when $k=k_n$, $n=1,2,\ldots$.

\begin{theorem}
\label{thm:3}
 Let the components of $\vec{L}$ be linearly independent over ${\mathbb Q}$.
Then there exists a probability density $P_v(y)$, depending
on $\vec{L}$, such that
\begin{equation*}
  \lim_{N\to\infty}\frac{1}{N}{\#}\!\left\{n\in\{1,\ldots,N\} : \frac{1}{v^2}
Z'(k_n,\vec{L})<R\right\}=\int_{-\infty}^{R} P_v(y)\rmd y,
\end{equation*}
with $P_v(y)=0$ for $y<0$.
\end{theorem}
\begin{theorem}
\label{thm:4}
For each $v$ let the bond lengths $L_j$, $j=1,\ldots,v$ lie in the range 
$[\Lbar,\Lbar+\Delta L]$ and be linearly independent
over ${\mathbb Q}$. If $v\Delta L\to0$ as $v\to\infty$
then for any $R\in{\mathbb R}$,
\begin{equation*}
  \int_{-\infty}^R P_v(y)\rmd y\to\int_{-\infty}^R P(y)\rmd y
\end{equation*}
as $v\to\infty$. The limiting density is given by the continuous function
  \begin{equation*}
  P(y)=\left\{
\begin{array}{lr}
\displaystyle \frac{\sqrt{\Lbar}}{4\pi y^{3/2}}\int_{-\infty}^{\infty}
\exp\left(-\frac{\xi^2}{4}-\frac{\Lbar m(\xi)^2}{4y}\right)m(\xi)
\rmd\xi,
&\mbox{$y>0$}\\
0, & \mbox{$y\leq0$},
\end{array}
\right.
\end{equation*}
where
\begin{equation*}
  m(\xi):=\frac{2}{\sqrt{\pi}}\rme^{-\xi^2/4}+\xi\erf(\xi/2).
\end{equation*}
\end{theorem}
By comparison, the value distribution for the 
logarithm of the derivative of the 
characteristic polynomial of a matrix drawn from the CUE of random
matrix theory, evaluated at an eigenvalue in the limit as matrix size 
tends to infinity, is Gaussian 
\cite{hug:rmt}.

The following results refer to the value distribution of $A_i(n,\vec{L};v)$.
\begin{theorem}
\label{thm:5}
  Assume the conditions of theorem \ref{thm:3} are satisfied. Then there
exists a probability density $Q_v(\eta)$ such that
\begin{equation*}
  \lim_{N\to\infty}\frac{1}{N}{\#}\!\left\{ n\in\{1,\ldots,N\}: v^2A_i
(n,\vec{L};v)<R\right\}=\int_0^{R} Q_v(\eta)\rmd\eta
\end{equation*}
where the density $Q_v(\eta)$ is independent of the choice of bond $i$ but
depends on $\vec{L}$.
\end{theorem}

\begin{theorem}
\label{thm:6}
Assume the conditions of theorem \ref{thm:4} are satisfied. Then for each $R>0$
\begin{equation*}
  \int_{0}^R Q_v(\eta)\rmd\eta\to\int_0^R Q(\eta)\rmd\eta.
\end{equation*}
as $v\to\infty$. The limiting density is given by the function
\begin{equation}
\label{eq:Q:def}
  Q(\eta)=\frac{1}{2\pi^{3/2}\eta}\Im
\int_{-\infty}^{\infty}\exp\left(-\frac{\xi^2}{4}-\frac{\Lbar \eta m(\xi)^2}
{8}\right)\erfc\left( \frac{\sqrt{\Lbar\eta}m(\xi)}{2\rmi
\sqrt{2}}\right)\rmd\xi
\end{equation}
which is continuous on $(0,\infty)$. 
Here $m(\xi)$ is as in theorem \ref{thm:4}.
$Q(\eta)$ has asymptotic expansion
\begin{equation} \label{eq:Q:asympt}
   Q(\eta)=\frac{\sqrt{2}}{\sqrt{\Lbar}\pi^2\eta^{3/2}}\int_{-\infty}^{\infty}
\frac{\rme^{-\xi^2/4}}{m(\xi)}\rmd\xi
+\Or(\eta^{-5/2})\qquad
\end{equation}
as $\eta\to\infty$.
\end{theorem}

The proofs of theorems \ref{thm:3}--\ref{thm:6} rely on an equidistribution
result of Barra and Gaspard \cite{bar:lsd}. We review this work in section
\ref{sec:3}.

The limit $v\to\infty$ is analogous to the semiclassical limit $\hbar\to 0$
\cite{kot:pot}. We note that theorem \ref{thm:6} describes the wave 
functions on a vanishingly small fraction of 
the graph. 
It thus goes beyond the information provided by the Schnirelman theorem.
It instead corresponds to the Gaussian value distribution for the wave
functions of classically chaotic systems implied by the random wave 
model \cite{ber:rwm}.

The value distribution of the eigenvector components of
asymptotically large random matrices are  
particular cases of the $\chi^2_{\beta}$ density
\begin{equation*}
P_{\chi_{\beta}^2}(\eta)=\left(\frac{\beta}{2}\right)^{\beta/2}
\eta^{\beta/2-1}\Gamma^{-1}\left(\frac{\beta}{2}\right)\rme^{-\beta\eta/2},
\end{equation*}
where the parameter $\beta$ takes the values $1$, $2$ and $4$ in, respectively,
the orthogonal, unitary and symplectic ensembles 
(see for example \cite{haa:rmt}). When
$\beta=1$ the density is called the Porter-Thomas density. It is characterised
by $\Or(\eta^{-1/2})$ behaviour as $\eta\to0$ and 
$\Or(\eta^{-1/2}\rme^{-\eta/2})$ as $\eta\to\infty$. The limiting distribution
we find in theorem \ref{thm:6} completely determines the value distribution
of the star graph eigenfunctions (see the appendix) and has a 
significantly different shape (c.f.\ equation
\eref{eq:Q:asympt} and figure \ref{fig:9:5} below). 
Other quantum systems for which the value distribution
of the eigenfunctions has a non-random-matrix limit are the Cat Maps 
\cite{par:cat}.

In \cite{berk:2} a correspondence was noted between the two-point spectral
correlation functions for star graphs and a class of systems known
as \v{S}eba billiards. The original \v{S}eba billiard \cite{seba:1} was
a rectangular quantum billiard perturbed by a point singularity.
More generally, we describe any integrable system perturbed in such a way as
belonging to the same class \cite{seba:2}. We conjecture that the results 
derived in the present work will also apply to systems in the \v{S}eba
class.

The remainder of this paper is structured as follows. In section \ref{sec:2}
we prove theorems \ref{thm:1} and \ref{thm:2}. In section \ref{sec:3} we
treat the finite $v$ cases, theorem \ref{thm:3} and \ref{thm:5}. In
sections \ref{sec:4} and \ref{sec:5} we prove, respectively, theorems
\ref{thm:4} and \ref{thm:6}, developing the necessary machinery in
section \ref{sec:4}. Section \ref{sec:6} is devoted to numerical computations
that illustrate our results. We develop more fully the connections
between the present work and \v{S}eba billiards in section \ref{sec:7}.

\section{The value distribution of $Z(\vec{L},k)$}
\label{sec:2}
\begin{lemma}
\label{lem:1}
Let $\Lbar>0$ and $\zeta$ be real constants, then 
\begin{equation*}
\lim_{k\to\infty}\frac{1}{\Delta L}\Lint \exp(\rmi\zeta\tan kL)\rmd L
=\rme^{-|\zeta|}
\end{equation*}
uniformly for $k\Delta L\to\infty$ as $k\to\infty$.
\end{lemma}
\dimonstrazione
By the periodicity of the integrand we may shift the range of integration
by multiples of $\pi/k$ so that without loss of generality we may take
$\Lbar$ in the range $0\leq\Lbar\leq \pi/k$. We write $\Delta L=\pi(n+p)/k$
where $n\in{\mathbb Z}$ and $0\leq p < 1$. Then by the periodicity of the
integrand,
\begin{align*}
\int_{\Lbar}^{\Lbar+\pi n/k+\pi p/k}\exp&(\rmi \zeta\tan kL)\rmd L \\
&= \int_{\Lbar}^{\Lbar+p\pi/k}\exp(\rmi\zeta\tan kL)\rmd L
+n\int_0^{\pi/k}\exp(\rmi\zeta\tan{kL})\rmd L 
\\
&= \int_{\Lbar}^{\Lbar+p\pi/k}\exp(\rmi\zeta\tan kL)\rmd L
+\frac{n}{k}\int_{-\infty}^{\infty} 
\frac{\rme^{\rmi\zeta z}}{1+z^2}\rmd z
\end{align*}
where the substitution $z=\tan kL$ has been made. We note now that
$n/k\Delta L\to\pi^{-1}$ as $k\to\infty$ and
\begin{equation*}
\left|\frac{1}{\Delta L}\int_{\Lbar}^{\Lbar+p\pi/k}\exp(\rmi\zeta\tan{kL})
\rmd L\right| \leq \frac{\pi}{k\Delta L} \to 0\quad \mbox{as $k\to\infty$.}
\end{equation*}
A simple application of Cauchy's residue theorem allows us to evaluate 
the integral 
\begin{equation}
\int_{-\infty}^{\infty}\frac{\rme^{\rmi\zeta z}}{1+z^2}\rmd z=
\pi\rme^{-|\zeta|}.
\label{eqn:10}
\end{equation}
\finire
\dimon{Proof of theorem \ref{thm:1}.}
We use here the characteristic function. With bond lengths chosen from a
uniform distribution,
\begin{eqnarray*}
{\mathbb E}_{\rm L}(\exp(\rmi\zeta Z(k,\vec{L})))&=&\frac{1}{(\Delta L)^v}
\Lint\mspace{-20mu}
\cdots\!\Lint\mspace{-15mu}\exp\left(\rmi\zeta\textstyle\sum_{j=1}^v 
\tan kL_j\right)\rmd L_1\cdots\rmd L_v \nonumber \\
&=&\left(\frac{1}{\Delta L}\Lint \exp(\rmi\zeta\tan kL)\rmd L
\right)^v.
\end{eqnarray*}
The subscript ${\rm L}$ indicates that the expectation is with respect to
an average over bond lengths.
Since the map $t\mapsto t^v$ is continuous, lemma \ref{lem:1} together 
with \eref{eqn:10} allows us to deduce that 
\begin{equation}
\lim_{k\to\infty}{\mathbb E}_{\rm L}(\exp(\rmi\zeta Z(k,\vec{L})))=
\rme^{-v|\zeta|}.
\label{eqn:11}
\end{equation}
The limiting density corresponding to the characteristic function on the
right hand side is given by 
\begin{equation*}
P_Z(x)=\frac{1}{2\pi}\int_{-\infty}^{\infty}\exp(-\rmi\zeta x-v|\zeta|)
\rmd\zeta =\frac{1}{\pi}\frac{v}{v^2+x^2}.
\end{equation*}
The theorem follows now from the
classical continuity theorem for characteristic functions (\cite{fel:ipt}
chapter XV).
\finire
The proof of theorem \ref{thm:2} uses Weyl's Equidistribution theorem.
This celebrated result \cite{wey:ube} has numerous applications in analysis
and number theory. We state here the form most convenient for application
to our current work.
Let ${\mathbb T}^v$ be the $v$-dimensional torus, with sides of
length $\pi$.
\begin{theorem}
\label{thm:weyl}
Let $f\in C({\mathbb T}^v)$, and let the components of $\vec{L}$
be linearly independent over ${\mathbb Q}$. Then
\begin{equation*}
\lim_{K\to\infty}\frac{1}{K}\int_0^K f(L_1k,\ldots,L_vk)\rmd k= 
\frac{1}{\pi^v}\int_{{\mathbb T}^v} f(\vec{x})\rmd \vec{x}
\end{equation*}
where $\displaystyle \rmd\vec{x}=\rmd x_1\cdots\rmd x_v$ denotes Lebesgue
measure. 
\end{theorem}
We shall use Weyl's theorem as our main tool to relate $k$-averages to bond
length averages. It is for this reason that it is crucial that the 
bond lengths are incommensurate.

We remark that theorem \ref{thm:weyl} can also apply to more general functions
such as piecewise continuous functions through an argument similar to the
one in the following lemma.
\begin{lemma}
\label{lem:2}
Theorem \ref{thm:weyl} can also be applied to the function 
\begin{equation*}
f(\vec{x}):=\exp\left(\textstyle \rmi\zeta\sum_{j=1}^v\tan x_j\right).
\end{equation*}
\end{lemma}
\dimonstrazione
We treat the real and imaginary parts of $f$ separately. The functions
\begin{eqnarray*}
f_1(\vec{x})&:=\cos\left(\textstyle \zeta\sum_{j=1}^{v}\tan x_j\right)\\
f_2(\vec{x})&:=\sin\left(\textstyle \zeta\sum_{j=1}^{v}\tan x_j\right)
\end{eqnarray*}
are smooth everywhere apart from at an essential singularity
when $x_i=\pi/2$ for some $i$, which we tame in the following way.
Let $\epsilon>0$ . We can construct functions $\phi$ and $\psi$ satisfying the 
conditions
of theorem \ref{thm:weyl} such that
\begin{eqnarray}
\psi(\vec{x})=-1 & &\mbox{if $|x_i-\pi/2|<\pi\epsilon^{1/v}/8$ for some 
$i=1,\ldots,v$,} \nonumber\\
\phi(\vec{x})=1 & &\mbox{if $|x_i-\pi/2|<\pi\epsilon^{1/v}/8$ for some 
$i=1,\ldots,v$,} \nonumber\\
\psi(\vec{x})=\phi(\vec{x})=f_1(\vec{x})& &\parbox{2.5in}{if 
$\pi\epsilon^{1/v}/4<|x_i-\pi/2|<\pi/2$ for some
$i=1,\ldots,v$,} \nonumber\\
-1\leq\psi(\vec{x})\leq f_1(\vec{x})\leq\phi(\vec{x})\leq 1
& &\mbox{for all $\vec{x}\in{\mathbb T}^v$.}
\label{eqn:13}
\end{eqnarray}
This implies
\begin{equation*}
\frac{1}{\pi^v}\int_{{\mathbb T}^v} \left(\phi(\vec{x})-\psi(\vec{x})
\right)\rmd\vec{x}
\leq \frac{1}{\pi^v}\left(2\frac{\pi}{2}\epsilon^{1/v}\right)^v =\epsilon.
\end{equation*}
From \eref{eqn:13} and theorem \ref{thm:weyl},
\begin{align*}
\frac{1}{\pi^v}\int_{{\mathbb T}^v} \psi(\vec{x})\rmd\vec{x} \leq
\liminf_{K\to\infty}&\frac{1}{K}\int_0^K f_1(k\vec{L})\rmd k \\
&\leq\limsup_{K\to\infty}\frac{1}{K}\int_0^K f_1(k\vec{L})\rmd k \leq
\frac{1}{\pi^v}\int_{{\mathbb T}^v} \phi(\vec{x})\rmd\vec{x}.
\end{align*}
The ends of this inequality differ by $\epsilon$ which can be made arbitrarily
small, so we see that $\lim_{K\to\infty} K^{-1}\int_0^K
f_1(k\vec{L})\rmd k$ exists and is equal to
\begin{equation*}
\frac{1}{\pi^v}\int_{{\mathbb T}^v}
f_1(\vec{x})\rmd\vec{x}.
\end{equation*}
The extension to $f_2$ and hence $f$ is obvious.
\finire

\dimon{Proof of theorem \ref{thm:2}.}
We begin in the same way as in the proof of theorem \ref{thm:1}. In this case
$k$ is chosen uniformly from the interval $[0,K]$ with $K>0$, so
that the characteristic function with respect to this uniform distribution
is
\begin{equation*}
{\mathbb E}_K(\exp(\rmi\zeta Z(k,\vec{L})))=\frac{1}{K}\int_0^K \exp\left(\rmi
\zeta \textstyle \sum_{j=1}^v \tan kL_j\right) \rmd k.
\end{equation*}
By lemma \ref{lem:2} we can write this integral as an average over the 
torus as $K\to\infty$:
\begin{eqnarray*}
\lim_{K\to\infty}{\mathbb E}_K(\exp(\rmi\zeta Z(k,\vec{L})))&=&
\frac{1}{\pi^v}
\int_{{\mathbb T}^v}\exp\left( \textstyle\rmi\zeta\sum_{j=1}^v \tan x_j\right)
\rmd\vec{x}\\
&=&\left( \frac{1}{\pi}\int_0^{\pi} \exp(\rmi\zeta\tan x)\rmd x \right)^v.
\end{eqnarray*}
Following the substitution $z=\tan x$ in this final integral, we have
\begin{equation*}
\lim_{K\to\infty}{\mathbb E}_K(\exp(\rmi\zeta Z(k,\vec{L}))=\rme^{-v|\zeta|}
\end{equation*}
and the theorem follows from the same arguments used in the end
of the proof of theorem \ref{thm:1}.
\finire
\section{An equidistribution theorem}
\label{sec:3}
Barra and Gaspard \cite{bar:lsd} observed that the condition for $k$ to
be an eigenvalue of a graph can be written in the form
\begin{equation*}
  G(k\vec{L})=0,
\end{equation*}
where $G$ is a function that is periodic in each variable. For star graphs 
$G$ is defined on ${\mathbb T}^v$ by
\begin{equation*}
  G(\vec{x})=\tan x_1+\cdots+\tan x_v.
\end{equation*}
The equation 
\begin{equation*}
  G(\vec{x})=0
\end{equation*}
defines a surface $\Sigma$ embedded in ${\mathbb T}^v$. A
flow $\phi^k, k\in{\mathbb R}$ can be defined on ${\mathbb T}^v$ by
\begin{equation}
  \phi^{k}(\vec{x}_0)=\vec{x}_0+k\vec{L} \quad\mbox{(mod $\pi$)}.
\end{equation}

Since $k=0$ is an eigenvalue for star graphs with Neumann boundary 
conditions considered here, we take $\vec{x}_0=0$ in this case.

At each value $k=k_n$ we have an intersection
of this flow with the surface $\Sigma$. 
We note that the angle between the normal to the surface $\Sigma$ and
the flow $\phi^k$ is given by
\begin{equation*}
  \cos\theta=\frac{|\vec{L}\cdot\nabla G|}{\|\vec{L}\|\|\nabla G\|}.
\end{equation*}
For star graphs,
\begin{equation*}
  \nabla G(\vec{x})=(\sec^2 x_1,\ldots,\sec^2 x_v).
\end{equation*}
Hence there exists a constant $c_1>0$ such that $\cos\theta> c_1$. This
means that the angle between the flow and the surface $\Sigma$ is uniformly
bounded away from 0. We can therefore parameterise
$\Sigma$ locally by $v-1$ real variables $\vecxi=(\xi_1,\ldots,\xi_{v-1})$
so that for $\vec{x}\in\Sigma$,
\begin{equation*}
  x_i=s_i(\vecxi)
\end{equation*}
and $G(s_1(\vecxi),\ldots,s_v(\vecxi))=0$.

The central result of Barra and Gaspard is the existence of
an invariant measure on the surface $\Sigma$.
\begin{theorem}
\label{thm:bar:gas}
Let $f$ be a piecewise continuous function $\Sigma\to{\mathbb R}$. Then
\begin{equation}
  \lim_{N\to\infty}\frac{1}{N}\sum_{n=1}^{N}f(k_n\vec{L})=\int_{\Sigma} 
f(\vecxi)\rmd\nu(\vecxi)
\label{eq:bar:gas}
\end{equation}
where the measure $\nu$ is given by
\begin{equation}
  \rmd\nu(\vecxi)=\frac{J(\vecxi)\rmd\vecxi}
{\int_{\Sigma}J(\vecxi)\rmd\vecxi}
\end{equation}
and $\rmd\vecxi=\rmd\xi_1\cdots\rmd\xi_{v-1}$ is Lebesgue measure.
$J$ is the Jacobian determinant
\begin{equation*}
  J(\vecxi)=\left|
  \begin{array}{ccc}
L_1 & \cdots & L_v \\
\frac{\partial s_1}{\partial \xi_1} & \cdots &
\frac{\partial s_v}{\partial \xi_1}\\
\vdots & \ddots & \vdots  \\
\frac{\partial s_1}{\partial \xi_{v-1}} & \cdots &
\frac{\partial s_v}{\partial \xi_{v-1}}
\end{array} \right|
\end{equation*}
\end{theorem}

For completeness, we sketch a proof of theorem \ref{thm:bar:gas} 
for star graphs with $v$ bonds. 

\dimonstrazione
  Let $\tilde{f}:{\mathbb T}^v\to{\mathbb R}$ be an extension of $f$ 
to ${\mathbb T}^v$, so that $\tilde{f}\big|_{\Sigma}=f$, i.e.\
\begin{equation*}
  \tilde{f}(\vec{x})=f(\vec{x})\qquad\mbox{for all $\vec{x}\in\Sigma$.}
\end{equation*}
We let $\tilde{f}$ be constructed in such a way that
for all $\vecxi\in\Sigma$, 
$\tilde{f}(\phi^{k}(\vecxi))$ is a differentiable function of $k$ with 
compact support in some neighbourhood of $k=0$.

Let $\epsilon>0$.
We construct the set $\Sigma_{\epsilon,\vec{L}}$
which is a thickening of $\Sigma$ in the direction of the flow $\phi^k$,
\begin{equation*}
  \Sigma_{\epsilon,\vec{L}}:=\{\vec{x}\in{\mathbb T}^v :\exists\vecxi\in
\Sigma, k\in[-\epsilon,\epsilon]:\vec{x}=\phi^k(\vecxi)\}\subseteq
{\mathbb T}^v.
\end{equation*}
We define $\I_A$, the indicator function of a set $A$, by
\begin{equation*}
  \I_A(x):=\left\{
  \begin{array}{lr}
    1, & x\in A\\
    0, & x\not\in A
  \end{array}\right.
\end{equation*}
The indicator function $\I_{\Sigma_{\epsilon,\vec{L}}}(\vec{x})$ is piecewise constant.

By the differentiability properties of $\tilde{f}$ we can write
for every $\vec{x}\in\Sigma$
\begin{equation}
  f(\vec{x})=\tilde{f}(\vec{x})=
\frac{1}{2\epsilon}\int_{-\epsilon}^{\epsilon}
\tilde{f}(k\vec{L}+\vec{x})\rmd k+\Or(\epsilon)
\end{equation}
as $\epsilon\to 0$. The implied constant does not depend on $\vec{x}$. Setting
$\vec{x}=k_n\vec{L}$ gives
\begin{equation*}
  \frac{1}{N}\sum_{n=1}^N f(k_n\vec{L})=\frac{1}{2\epsilon N}\sum_{n=1}^N
\int_{-\epsilon}^{\epsilon}\tilde{f}((k+k_n)\vec{L})\rmd k+\Or(\epsilon).
\end{equation*}
Let the mean density of zeros of $Z(k,\vec{L})$ be $\bar{d}$:
\begin{equation}
  \bar{d}:=\lim_{K\to\infty}\frac{\#\{n:k_n\leq K\}}{K}.
\end{equation}
Then
\begin{eqnarray*}
  \lim_{N\to\infty}\frac{1}{N}\sum_{n=1}^N f(k_n\vec{L})&=&\frac{1}
{2\epsilon\bar{d}}\lim_{K\to\infty}\frac{1}{K}\int_0^{K} \tilde{f}(k\vec{L})
\I_{\Sigma_{\epsilon,\vec{L}}}(k\vec{L})\rmd k+\Or(\epsilon) \\
&=&\frac{1}{\pi^v \bar{d}}\int_{{\mathbb T}^v} \tilde{f}(\vec{x})
\I_{\Sigma_{\epsilon,\vec{L}}}(\vec{x})\rmd\vec{x} +\Or(\epsilon),
\end{eqnarray*}
applying theorem \ref{thm:weyl} to the piecewise continuous function 
$\tilde{f}(\vec{x})
\I_{\Sigma_{\epsilon,\vec{L}}}(\vec{x})$. 
Changing to the system of coordinates $(t,\vecxi)$
on ${\mathbb T}^v$ via the change of variables
\begin{equation*}
  x_i=L_i t+s_i(\vecxi),
\end{equation*}
gives
\begin{equation*}
\lim_{N\to\infty}\frac{1}{N}\sum_{n=1}^N f(k_n\vec{L})=\frac{1}
{\pi^v\bar{d}}\int_{\Sigma}\frac{1}{2\epsilon}\int_{-\epsilon}^{\epsilon} 
\tilde{f}(t,\vecxi)J(\vecxi)\rmd t\rmd\vecxi+\Or(\epsilon).
\end{equation*}
Since this is true for all $\epsilon>0$, we deduce that
\begin{equation}
  \lim_{N\to\infty}\frac{1}{N}\sum_{n=1}^N f(k_n\vec{L})=\frac{1}
{\pi^v\bar{d}}\int_{\Sigma}{f}(\vecxi)J(\vecxi)\rmd\vecxi.
\end{equation}
By setting $f=1$, we see that
\begin{equation}
  \bar{d}=\frac{1}{\pi^v}\int_{\Sigma}J(\vecxi)\rmd\vecxi,
\end{equation}
to complete the proof.
\finire
We note incidentally that $\bar{d}$ can be evaluated using spectral methods
\cite{kot:pot} to give
\begin{equation*}
  \bar{d}=\frac{1}{\pi}\sum_{j=1}^v L_j=\frac{v\Lbar}{\pi}+\Or(v\Delta L)
\qquad\mbox{as $v\Delta L\to 0$.}
\end{equation*}

We observe that the right hand side of equation \eref{eq:bar:gas} can formally 
be written in the form
\begin{equation}
\int_{\Sigma} 
f(\vecxi)\rmd\nu(\vecxi)
  =\frac{1}{2\pi^{v+1}
\bar{d}}\int_{-\infty}^{\infty}
\int_0^{\pi}\!\cdots\!\int_0^{\pi} f(\vec{x})[\vec{L}\cdot\nabla 
G(\vec{x})]\rme^{\rmi\zeta G}\rmd x_1\cdots\rmd x_v\rmd\zeta,
\label{eq:states}
\end{equation}
where now $f$ is a function ${\mathbb T}^v\to{\mathbb R}$ of an appropriate
class.
This follows from writing $\int_{\Sigma} 
f(\vecxi)\rmd\nu(\vecxi)$ in the equivalent form,
\begin{equation}
  \frac{1}{\pi^v\bar{d}}
\int_{\Sigma}\int f(t,\vecxi)[\vec{L}\cdot\nabla G] \delta(G(t,\vecxi))
J(\vecxi)\rmd t\rmd
\vecxi.
\label{eq:del:seq}
\end{equation}
We then write the $\delta$-function as the limit of the sequence
\begin{equation}
\delta_{m}(x):=\frac{1}{2\pi}\int_{-m}^{m}\rme^{\rmi\zeta x}\rmd\zeta=
\frac{\sin mx}{\pi x}
\label{eq:delta}
\end{equation}
as $m\to\infty$, and changing back to the usual Cartesian coordinates on
${\mathbb T}^v$. We need to show that $\delta_m$ is an appropriate 
$\delta$-sequence for the function $f$ that we consider.
We will check this point directly in the 
calculations where the identity \eref{eq:states} is used. We shall also justify
taking the $\zeta$-integral in \eref{eq:delta} outside the integral 
over ${\mathbb T}^v$.

\begin{proposition}
  For a star graph with $v$ bonds and the parameterisation
\begin{eqnarray*}
  s_i=\xi_i\qquad\mbox{$i=1,\ldots,v-1$}\\
s_v=-\tan^{-1}(\tan \xi_1+\cdots+\tan\xi_{v-1})
\end{eqnarray*}
$J(\vecxi)$ takes the following form
  \begin{equation}
 J(\vecxi)=\frac{L_1\sec^2\xi_1+\cdots+L_{v-1}\sec^2{\xi_{v-1}}}
{1+(\tan \xi_1+\cdots+\tan \xi_{v-1})^2} +L_v
\label{eqn:j}
  \end{equation}
\end{proposition}
\dimonstrazione
Differentiating gives
\begin{equation*}
  \frac{\partial s_i}{\partial \xi_j}=\delta_{ij}
\end{equation*}
for $i<v$ and
\begin{equation*}
  \frac{\partial s_v}{\partial \xi_j}=\frac{-\sec^2{\xi_j}}{1+(\tan\xi_1+
\cdots+\tan\xi_{v-1})^2}.
\end{equation*}
For ease of notation we write $D:={1+(\tan\xi_1+
\cdots+\tan\xi_{v-1})^2}$.
Thus we have
\begin{equation*}
  J(\vecxi)=\left|
  \begin{array}{ccccc}
L_1 & L_2 & \cdots & L_{v-1} & L_v \\
1   &  0  & \cdots & 0 & \frac{-\sec^2\xi_1}{D} \\
0   &  1  & \cdots & 0 & \frac{-\sec^2\xi_2}{D} \\
\vdots & \vdots & \ddots & \vdots & \vdots \\
0 & 0 & \cdots & 1 & \frac{-\sec^2\xi_{v-1}}{D}
  \end{array}\right|.
\end{equation*}
To complete the proof we employ the identity
\begin{equation*}
\left|  \begin{array}{ccccc}
\alpha_1 & \alpha_2 & \cdots & \alpha_{n-1} & \alpha_n \\
1   &  0  & \cdots & 0 & \beta_1 \\
0   &  1  & \cdots & 0 & \beta_2 \\
\vdots & \vdots & \ddots & \vdots & \vdots \\
0 & 0 & \cdots & 1 & \beta_{n-1}
  \end{array}\right|
=(-1)^n\left( -\alpha_n+\sum_{k=1}^{n-1}\alpha_k\beta_k\right),
\end{equation*}
which may be readily checked by induction.
\finire
We note in passing that the explicit form of $J(\vecxi)$ given 
above together with theorem 
\ref{thm:bar:gas} provides a convenient representation for use numerical 
studies of eigenvalues and eigenfunctions because
the zeros of $Z(k,\vec{L})$ do not need to be computed explicitly.
\dimon{Proof of theorem \ref{thm:3}.}
We take as the function $f$ in theorem \ref{thm:bar:gas}
\begin{equation*}
f(\vec{x})=\I_{(-\infty,R]}\left(\frac{1}{v^2}\sum_{j=1}^vL_j\sec^2{x_j}\right)
\end{equation*}
Then we define $P_v(y)$ by
\begin{eqnarray*}
\int_{-\infty}^R P_v(y)\rmd y=\frac{1}{\pi^v\bar{d}}\int_0^{\pi}
\!\cdots\!\int_0^{\pi}f(\xi_1,\ldots,\xi_{v-1},-\tan^{-1}(\tan \xi_1+
\cdots+\tan \xi_{v-1}))\\ \times J(\vecxi)\rmd\xi_1\cdots\rmd\xi_{v-1},
\end{eqnarray*}
where $J(\vecxi)$ is defined by \eref{eqn:j}.
Since $\sec^2x>0$ for all $x\in{\mathbb R}$, it follows that 
$P_v(y)=0$ for $y<0$.
\finire
\dimon{Proof of theorem \ref{thm:5}.}
In this case, we take as the function $f$ in theorem \ref{thm:bar:gas},
\begin{equation*}
 f(\vec{x})=\I_{[0,R]}\left(\frac{2v^2\sec^2x_i}{\sum_j L_j\sec^2{x_j}}\right).
\end{equation*}
We take as the parameterisation of $\Sigma$
\begin{eqnarray*}
  s_j&=&\xi_j,\quad\mbox{$1\leq j<i$}\\
  s_i&=&-\tan^{-1}(\tan\xi_1+\cdots+\tan\xi_{v-1}) \\
  s_j&=&\xi_{j-1},\quad\mbox{$i<j\leq v$}.
\end{eqnarray*}
This does not change the form of $J(\vecxi)$ from that in \eref{eqn:j}, 
but introduces extra symmetry in $f$. Then define, as before,
\begin{equation*}
 \int_0^R P_v(\eta)\rmd\eta=\frac{1}{\pi^v\bar{d}}
\int_0^{\pi}\!\cdots\!\int_0^{\pi}f(s_1(\vecxi),\ldots,s_v(\vecxi))
J(\vecxi)\rmd\xi_1\cdots\rmd\xi_{v-1}.
\end{equation*}
Since $f(\vec{s}(\vecxi))$ is symmetric in $\xi_1,\ldots,\xi_{v-1}$ we
see that $P_v(\eta)$ is independent of the choice of bond $i$.
\finire

\section{Value distribution of $Z'(k_n)$ in the limit $v\to\infty$}
\label{sec:4}
We take as the function $f$ in \eref{eq:states} the characteristic function
for the value distribution of the derivative of $Z(k,\vec{L})$,
\begin{equation}
  Z'(k,\vec{L})=\sum_{j=1}^v L_j\sec^2 kL_j.
\label{eq:5:1}
\end{equation}
Since we are stipulating that $v\Delta L\to0$ as $v\to\infty$, we replace 
$L_j$ by $\Lbar$ where it does not
multiply $k$ in \eref{eq:5:1} and take as our $f$
\begin{equation*}
  f(\vec{x})=\exp\left(-\frac{\rmi\beta\Lbar}{v^2}\sum_{j} \sec^2x_j\right).
\end{equation*}
Let the quantity in which we are interested be denoted $E_v(\beta)$.
Then 
\begin{eqnarray}
  E_v(\beta)&:=&\lim_{N\to\infty}\frac{1}{N}\sum_{n=1}^{N}
\exp\left(-\frac{\rmi\beta\Lbar}{v^2}\sum_j \sec^2 k_nL_j\right) \nonumber\\
 &=&\frac{\Lbar}{2\pi\bar{d}v}\int_{-\infty}^{\infty}\frac{1}{\pi^v}
\int_0^{\pi}\!\cdots\!\int_0^{\pi}
\left(\sum_j \sec^2{x_j}\right) \nonumber \\
& &\times\prod_{j=1}^v\exp\left(-\frac{\rmi\beta\Lbar}{v^2}\sec^2 x_j+
\frac{\rmi\zeta}{v}\sum_j\tan{x_j}\right)
\rmd\vec{x}\rmd\zeta. \label{jon:23}
\end{eqnarray}
We have made the re-scaling $\zeta\mapsto\zeta/v$. This is a natural 
normalisation since $Z(k,\vec{L})$ is a sum of $v$ terms.

We exploit the
symmetry in the integral in \eref{jon:23} to write
\begin{equation*}
E_v(\beta)=\frac{1}{2v}\int_{-\infty}^{\infty} I_1(\beta,\zeta) 
(I_2(\beta,\zeta))^{v-1}\rmd\zeta,
\end{equation*}
where we have replaced $\bar{d}$ by $\Lbar v/\pi$ and defined the integrals
\begin{equation*}
I_1(\beta,\zeta):=\frac{1}{\pi}\int_0^{\pi}\sec^2{x}\exp\left(-
\frac{\rmi\beta\Lbar}{v^2}
\sec^2{x}+\frac{\rmi\zeta}{v}\tan x\right)\rmd x
\end{equation*}
and
\begin{equation*}
I_2(\beta,\zeta):=\frac{1}{\pi}\int_0^{\pi}\exp\left(-\frac{\rmi\beta\Lbar}
{v^2}\sec^2{x}+\frac{\rmi\zeta}{v}\tan x\right)\rmd x.
\end{equation*}
We note that $I_2$ is uniformly convergent in $\zeta$ but that $I_1$ is not.

\subsection{The integrals $I_1$ and $I_2$}
The substitution $z=\tan x$ gives
\begin{eqnarray}
  I_1(\beta,\zeta)&=&\frac{1}{\pi}\int_{-\infty}^{\infty} \exp\left(
-\frac{\rmi\beta\Lbar}{v^2}(1+z^2)+\frac{\rmi\zeta z}{v}\right)\rmd z \nonumber
\\
&=&\frac{v}{\sqrt{\pi}}\frac{1}{\sqrt{\rmi\beta\Lbar}}
\exp\left(-\frac{\zeta^2}{4\rmi\beta\Lbar} -\frac{\rmi\beta\Lbar}{v^2}\right)
\end{eqnarray}
quoting a standard integral. 

That $\delta_m$ defined in \eref{eq:delta} is a $\delta$-sequence for the
function $\exp(\rmi\alpha z^2)$ follows from the equality
\begin{equation*}
  \frac{1}{2\pi}\int_{-\infty}^{\infty}\int_{-\infty}^{\infty} \rme^{\rmi
\alpha z^2+\rmi\zeta(z-w)}\rmd z\rmd\zeta=\exp(\rmi\alpha w^2),
\end{equation*}
which can be checked by direct evaluation of the integrals. This and uniform 
convergence of $I_2$ in $\zeta$ justifies the operations that lead to
identity \eref{eq:states}.

We can treat $I_2$ in a similar manner to that in which we treated $I_1$.
\begin{equation*}
I_2(\beta,\zeta)=\frac{1}{\pi}
\int_{-\infty}^{\infty}\exp\left(-\frac{\rmi \beta\Lbar}{v^2}(1+z^2)+\frac{\rmi
\zeta z}{v}\right)\frac{\rmd z}{1+z^2}.
\end{equation*}
We first write 
\begin{equation*}
  \frac{1}{1+z^2}=\frac{1}{2\rmi}\left(\frac{1}{z-\rmi}-\frac{1}{z+\rmi}\right)
\end{equation*}
so that $I_2$ can be decomposed into a difference of two similar integrals
\begin{equation}
  I_2(\beta,\zeta)=:I^{-}_2(\beta,\zeta)-I^{+}_2(\beta,\zeta).
\end{equation}
Observing that
\begin{equation*}
  \exp\left(-\frac{\rmi\beta\Lbar}{v^2}(1+z^2)+\frac{\rmi\zeta z}{v}\right)=
\exp\left(-\frac{\rmi\beta\Lbar}{v^2}-\frac{\zeta^2}{4\rmi\beta\Lbar}-
\frac{\rmi\beta\Lbar}{v^2}\left(z+\frac{\zeta v}{2\beta\Lbar}\right)^2\right),
\end{equation*}
we can write
\begin{eqnarray*}
  I_2^{-}(\beta,\zeta)&:=&\frac{1}{2\pi\rmi}\int_{-\infty}^{\infty}\exp\left(
-\frac{\rmi\beta\Lbar}{v^2}(1+z^2)+\frac{\rmi\zeta z}{v}\right)\frac{\rmd z}
{z-i}\\
&=&\frac{1}{2\pi\rmi}\exp\left(-\frac{\rmi\beta\Lbar}{v^2}+\frac{\rmi\zeta^2}
{4\beta\Lbar}\right)
\int_{-\infty}^{\infty}\frac{\exp(-\rmi\beta\Lbar y^2/v^2)}
{y-\rmi-\zeta v/2\beta\Lbar}\rmd y
\end{eqnarray*}
via $y=z+\zeta v/2\beta\Lbar$. We make the change of variable
\begin{equation*}
  r=\frac{\sqrt{\rmi\beta\Lbar}}{v}y.
\end{equation*}
This is permitted, because it rotates the contour of integration into the 
second and fourth quadrants of the complex plane, where the analytic function 
$\rme^{-\rmi z^2}$ decays rapidly. 

\begin{figure}[h]
\begin{center}
\setlength{\unitlength}{7cm}
\begin{picture}(1,1)
\put(0,0){\includegraphics[angle=0,width=7.0cm,height=6cm]{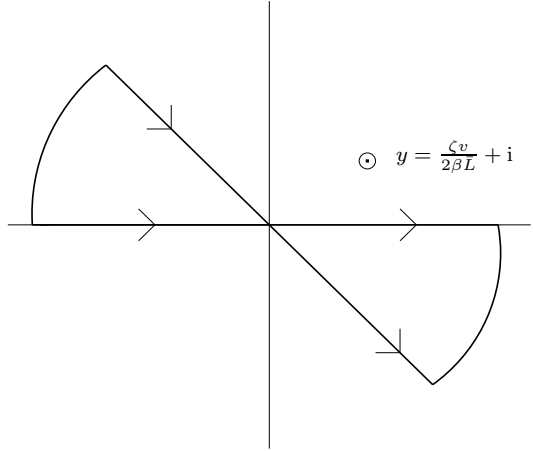}}
\put(0.74,0.55){$y=\frac{\zeta v}{2\beta\Lbar}+\rmi$}
\end{picture}
\caption{Deforming the contour of integration avoiding pole}
\label{fig:1}
\end{center}
\end{figure}

In the case $\zeta v/2\beta\Lbar > -1$ the new contour of 
integration avoids the pole at $y=\zeta v/2\beta\Lbar+\rmi$ 
(figure \ref{fig:1}) and Cauchy's
Theorem yields
\begin{equation*}
  I_2^{-}(\beta,\zeta)=
\frac{1}{2\pi\rmi}\exp\left(-\frac{\rmi\beta\Lbar}{v^2}+\frac{\rmi\zeta^2}
{4\beta\Lbar}\right)
\int_{-\infty}^{\infty}\frac{\rme^{-r^2}}{r-\frac{\sqrt{\rmi\beta\Lbar}}{v}(
\rmi+\frac{\zeta v}{2\beta\Lbar})}\rmd r.
\end{equation*}
This integral is standard, and may be found in, for example, \cite{abr:ste} 
(Equation {\bf 7.1.4}): 
\begin{equation*}
  \frac{\rmi}{\pi}\int_{-\infty}^{\infty} \frac{\rme^{-t^2}}{z-t} \rmd t
=\rme^{-z^2}\erfc(-\rmi z),\qquad\mbox{for $\Im\:z>0$.} 
\end{equation*}
The result we get is
\begin{eqnarray*}
  I^{-}_2(\beta,\zeta)&=&\frac{1}{2}\exp\left(-\frac{\rmi\beta\Lbar}{v^2}+
\frac{\rmi\zeta^2}{4\beta\Lbar}\right)\exp\left(-\frac{\rmi\beta\Lbar}{v^2}
\left(\rmi+\frac{\zeta v}{2\beta\Lbar}\right)^2\right)\\
& &\times\erfc\left(
\frac{\sqrt{\rmi\beta\Lbar}}{v}\left(1-\frac{\rmi\zeta v}{2\beta\Lbar}
\right)\right)\\
&=&\frac{1}{2}\exp\left(\frac{\zeta}{v}\right)\erfc\left(\frac{\zeta}{2\sqrt{
\rmi\beta\Lbar}}+\frac{\sqrt{\rmi\beta\Lbar}}{v}\right).
\end{eqnarray*}

\begin{figure}[h]
\begin{center}
\setlength{\unitlength}{7cm}
\begin{picture}(1,1)
\put(0,0){\includegraphics[angle=0,width=7.0cm,height=6cm]{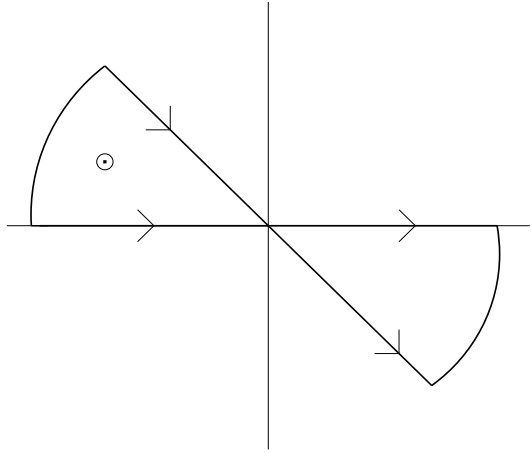}}
\end{picture}
\caption{Deforming the contour of integration enclosing pole}
\label{fig:2}
\end{center}
\end{figure}
If $\zeta v/2\beta\Lbar<-1$ then the contour encloses a pole 
(figure \ref{fig:2}). In this case,
\begin{equation*}
  \int_{-\infty}^{\infty}\frac{\exp(-\rmi\beta\Lbar y^2/v^2)}
{y-\rmi-\zeta v/2\beta\Lbar}\rmd y=2\pi\rmi R +\int_{-\infty}^{\infty}
\frac{\rme^{-r^2}}{r-\frac{\sqrt{\rmi\beta\Lbar}}{v}(\rmi+\frac{\zeta v}
{2\beta\Lbar})}\rmd r
\end{equation*}
where $R$ is the residue at the pole
\begin{eqnarray*}
  R&=&\left.\exp\left(-\frac{\rmi\beta\Lbar y^2}{v^2}\right)\right|_{y=\rmi+
\zeta v /2\beta\Lbar}\\
&=&\exp\left(-\frac{\rmi\zeta^2}{4\beta\Lbar}+\frac{\rmi\beta\Lbar}{v^2}+
\frac{\zeta}{v}\right),
\end{eqnarray*}
so that we also get in this case
\begin{equation*}
  I_2^{-}(\beta,\zeta)=\frac{1}{2}\exp\left(\frac{\zeta}{v}\right)
\erfc\left(\frac{\zeta}{2\sqrt{\rmi\beta\Lbar}}+\frac{\sqrt{\rmi\beta\Lbar}}
{v}\right).
\end{equation*}

Treating $I_2^{+}$ in a similar way, yields an expression for $I_2$,
\begin{equation}
I_2(\beta,\zeta)=\frac{1}{2}\rme^{\zeta/v}\erfc\left(
\frac{\zeta}{2\sqrt{\rmi\beta\Lbar}}+\frac{\sqrt{\rmi\beta\Lbar}}{v}\right)
+ \frac{1}{2}\rme^{-\zeta/v}\erfc\left(
\frac{-\zeta}{2\sqrt{\rmi\beta\Lbar}}+\frac{\sqrt{\rmi\beta\Lbar}}{v}\right).
\label{eq:5:2}
\end{equation}
\subsection{Some estimates}
\begin{lemma}
Let $-\sqrt{v}\leq\zeta\leq\sqrt{v}$, then
\begin{eqnarray}
  I_1(\beta,\zeta)(I_2(\beta,\zeta))^{v-1}=\frac{v}{\sqrt{\pi}}\frac{1}
{\sqrt{\rmi\beta\Lbar}}\exp\left(\frac{-\zeta^2}{4\rmi\beta\Lbar}\right)
M(\beta,\zeta)\nonumber \\
\times\left(1+\Or(v^{-1/2})+\Or(\zeta^2/v)\right)
\label{eq:5:30}
\end{eqnarray}
as $v\to\infty$, where
\begin{equation}
  M(\beta,\zeta):=\exp\left(-\frac{2\sqrt{\rmi\beta\Lbar}}{\sqrt{\pi}}
\exp\left(\frac{-\zeta^2}{4\rmi\beta\Lbar}\right)-
{\zeta}\erf\left(\frac{\zeta}{2\sqrt{\rmi\beta\Lbar}}\right) \right).
\end{equation}
\end{lemma}
\dimonstrazione
A key step will be to make a uniform expansion of $I_2(\beta,\zeta)$ as 
$v\to\infty$.
By Taylor's theorem, 
\begin{equation}
  \exp\left(\pm\frac{\zeta}{v}\right)=1\pm\frac{\zeta}{v}+\Or(
\zeta^2{v^{-2}}) \qquad\mbox{as $v\to\infty$.}\label{eq:5:17}
\end{equation}
A second application of Taylor's theorem yields
\begin{equation}
  \erfc\left(\frac{\pm\zeta}{2\sqrt{\rmi\beta\Lbar}}+
\frac{\sqrt{\rmi\beta\Lbar}}{v}\right)=\erfc\left(\frac{\pm\zeta}{2
\sqrt{\rmi\beta\Lbar}}\right)-\frac{2}{\sqrt{\pi}}\frac{\sqrt{\rmi\beta\Lbar}}
{v}\exp\left(-\frac{\zeta^2}{4\rmi\beta\Lbar}\right)+\frac{R_{1}(\zeta)}
{v^2},
\label{eq:erfc}
\end{equation}
where the remainder term is
\begin{equation*}
  R_1(\zeta)=-{\rmi\beta\Lbar}\int_0^1 \left.\frac{\rmd^2}
{\rmd z^2}\erfc(z)\right|_{z=\frac{\pm\zeta}{2\sqrt{\rmi\beta\Lbar}}+\sigma
\frac{\sqrt{\rmi\beta\Lbar}}{v}}(1-\sigma)\rmd\sigma.
\end{equation*}
The second derivative of $\erfc$ is
\begin{equation*}
  \frac{\rmd^2}{\rmd z^2}\erfc(z)=\frac{4}{\sqrt{\pi}}z\rme^{-z^2}.
\end{equation*}
This allows us to estimate
\begin{eqnarray}
  |R_1|&\leq&\frac{4\beta\Lbar}{\sqrt{\pi}}\int_0^1 \left(\frac{|\zeta|}
{2\sqrt{\beta\Lbar}}+\sigma\frac{|\sqrt{\beta\Lbar}|}{v}\right)
\rme^{\mp\sigma\zeta/v}(1-\sigma)\rmd\sigma \nonumber\\
&\leq&\frac{4\beta\Lbar}{\sqrt{\pi}}\rme^{|\zeta|/v}\left(\frac{|\zeta|}
{2\sqrt{\beta\Lbar}}+\Or(v^{-1})\right) \nonumber \\
&=&\Or(\sqrt{v}) \qquad\mbox{as $v\to\infty$,}\label{eq:5:21}
\end{eqnarray}
uniformly for $|\zeta|<\sqrt{v}$.
Substituting \eref{eq:5:17} and \eref{eq:erfc} into \eref{eq:5:2} gives
\begin{equation}
  I_2(\beta,\zeta)=1-\frac{2\sqrt{\rmi\beta\Lbar}}{v\sqrt{\pi}}
\exp\left(\frac{-\zeta^2}{4\rmi\beta\Lbar}\right)-
\frac{\zeta}{v}\erf\left(\frac{\zeta}{2\sqrt{\rmi\beta\Lbar}}\right)
+\frac{R_2(\zeta)}{v^2},
\label{eq:5:22}
\end{equation}
where $R_2$ is a combination of the errors in \eref{eq:5:17} and 
\eref{eq:5:21}. So
\begin{equation}
|R_2(\zeta)|=\Or(1+\zeta^2)\qquad\mbox{as $v\to\infty$.}
\label{eq:5:222}
\end{equation}

We thus have that 
\begin{align}
  \frac{1}{\sqrt{v}}\Bigg|\frac{2\sqrt{\rmi\beta\Lbar}}{\sqrt{\pi}}
&\exp\left(\frac{-\zeta^2}{4\rmi\beta\Lbar}\right)+
{\zeta}\erf\left(\frac{\zeta}{2\sqrt{\rmi\beta\Lbar}}\right)
-\frac{R_2(\zeta)}{v} \Bigg| \nonumber  \\
&\leq\frac{2\sqrt{\beta\Lbar}}{\sqrt{\pi v}}+\frac{|\zeta|}{\sqrt{v}}
+\frac{|R_2|}{v^{3/2}} \nonumber \\
&<2\qquad\mbox{for all $\zeta\in[-\sqrt{v},\sqrt{v}]$ and $v$ sufficiently 
large.}
\label{eq:5:6}
\end{align}

We are now in a position to understand the asymptotics of
$(I_2(\beta,\zeta))^{v-1}$. We first consider
\begin{equation}
  \ln\left(1-\frac{a}{v}\right)^{v-1}=(v-1)\left(-\frac{a}{v}
+\frac{R_3(a)}{v^2}\right), \label{eq:5:7}
\end{equation}
where 
\begin{equation*}
  |R_3(a)|\leq\frac{|a|^2}{1-|a|/v}
\end{equation*}
provided that $|a|<v$.
We take
\begin{equation*}
  a=\frac{2\sqrt{\rmi\beta\Lbar}}{\sqrt{\pi}}
\exp\left(\frac{-\zeta^2}{4\rmi\beta\Lbar}\right)+
{\zeta}\erf\left(\frac{\zeta}{2\sqrt{\rmi\beta\Lbar}}\right)
+\frac{R_2(\zeta)}{v}
\end{equation*}
which satisfies $|a|/\sqrt{v}<2$ for $v$ sufficiently large by \eref{eq:5:6}
and $a=\Or(|\zeta|)$ as $|\zeta|\to\infty$ for $|\zeta|<\sqrt{v}$ by
\eref{eq:5:222}. We see that
\begin{equation*}
\frac{|R_3(a)|}{v}=\Or(|a|^2/v)\qquad\mbox{as $v\to\infty$,}
\end{equation*}
so exponentiation of \eref{eq:5:7} gives
\begin{equation}
\left(1-\frac{a}{v}\right)^{v-1}=\rme^{-a}\left(1+\Or(v^{-1/2})\right)
\left(1+\Or((1+\zeta^2)/v)\right),
\label{eq:5:28}
\end{equation}
where the error estimates are uniform for $|\zeta|<\sqrt{v}$ as $v\to\infty$
and
\begin{equation*}
  \rme^{-a}=\exp\left(-\frac{2\sqrt{\rmi\beta\Lbar}}{\sqrt{\pi}}
\exp\left(\frac{-\zeta^2}{4\rmi\beta\Lbar}\right)-
{\zeta}\erf\left(\frac{\zeta}{2\sqrt{\rmi\beta\Lbar}}\right) \right)
\left[1+\Or((1+\zeta^2)/v)\right].
\end{equation*}
We can also write
\begin{equation}
  I_1(\beta,\zeta)=\frac{v}{\sqrt{\pi}}\frac{1}{\sqrt{\rmi\beta\Lbar}}
\exp\left(\frac{-\zeta^2}{4\rmi\beta\Lbar}\right)\left(1+\Or(v^{-2})\right).
\label{eq:5:29}
\end{equation}
Thus, taking \eref{eq:5:29} together with \eref{eq:5:22} and \eref{eq:5:28}
gives the required estimate.
\finire
\begin{lemma} \label{lem:5:7}
  Let $|\zeta|>\sqrt{v}$, then
  \begin{equation*} 
  |I_1(\beta,\zeta)(I_2(\beta,\zeta))^{v-1}|\leq\frac{1}{\pi}\left(
\frac{\beta\Lbar}{v^2}+\frac{\zeta^2}{\beta\Lbar}\right)^{-1}+\Or(v\zeta^{-3})
\qquad\mbox{as $|\zeta|\to\infty$.}
  \end{equation*}
\end{lemma}
\dimonstrazione
  We make use of the asymptotic expression
\begin{equation*}
  \erfc(z)=\frac{1}{\sqrt{\pi}}\rme^{-z^2}\left(\frac{1}{z}+\Or(z^{-3})\right)
\end{equation*}
as $z\to\infty$, valid for $|\arg z|<3\pi/4$. We see that
\begin{eqnarray*}
\rme^{\zeta/v}\erfc\left(\frac{\zeta}{2\sqrt{\rmi\beta\Lbar}}+\frac{
\sqrt{\rmi\beta\Lbar}}{v}\right)=\frac{1}{\sqrt{\pi}}\exp\left(\frac{-\zeta^2}
{4\rmi\beta\Lbar}+\frac{\rmi\beta\Lbar}{v^2}\right) \\
\times\left[\left(
\frac{\zeta}{2\sqrt{\rmi\beta\Lbar}}+\frac{
\sqrt{\rmi\beta\Lbar}}{v}\right)^{-1}+\Or(\zeta^{-3})\right].
\end{eqnarray*}
Similarly,
\begin{eqnarray*}
\rme^{-\zeta/v}\erfc\left(\frac{-\zeta}{2\sqrt{\rmi\beta\Lbar}}+\frac{
\sqrt{\rmi\beta\Lbar}}{v}\right)=\frac{1}{\sqrt{\pi}}\exp\left(\frac{-\zeta^2}
{4\rmi\beta\Lbar}+\frac{\rmi\beta\Lbar}{v^2}\right) \\
\times\left[\left(
\frac{-\zeta}{2\sqrt{\rmi\beta\Lbar}}+\frac{
\sqrt{\rmi\beta\Lbar}}{v}\right)^{-1}+\Or(\zeta^{-3})\right].
\end{eqnarray*}
Adding these gives the estimate
\begin{equation}
  |I_2(\beta,\zeta)|\leq\frac{\sqrt{\beta\Lbar}}{v\sqrt{\pi}}\left(
\frac{\beta\Lbar}{v^2}+\frac{\zeta^2}{\beta\Lbar}\right)^{-1}
+\Or(\zeta^{-3})\qquad\mbox{as $|\zeta|\to\infty$.}
\label{eq:5:3}
\end{equation}
Taking \eref{eq:5:3} together with the estimates $|I_2(\beta,\zeta)|\leq 1$
and
\begin{equation*}
  |I_1(\beta,\zeta)|\leq\frac{v}{\sqrt{\pi\beta\Lbar}}
\end{equation*}
gives us the estimate we require.
\finire
\begin{proposition}\label{prop:5:9}
  With the notation above,
  \begin{equation*}
   \lim_{v\to\infty}\frac{1}{2v}
\int_{-\infty}^{\infty} I_1(\beta,\zeta) (I_2(\beta,\zeta))^{v-1}
\rmd\zeta=\frac{1}{2\sqrt{\pi}}\int_{-\infty}^{\infty}
\frac{1}
{\sqrt{\rmi\beta\Lbar}}\exp\left(\frac{-\zeta^2}{4\rmi\beta\Lbar}\right)
M(\beta,\zeta)\rmd\zeta.
  \end{equation*}
\end{proposition}
\dimonstrazione
We split the region of integration as follows
\begin{eqnarray}
\frac{1}{2v}\int_{-\infty}^{\infty} I_1(\beta,\zeta)(I_2(\beta,\zeta))^{v-1}
\rmd\zeta =  \frac{1}{2v}\int_{-\sqrt{v}}^{\sqrt{v}} I_1(\beta,\zeta) 
(I_2(\beta,\zeta))^{v-1} \rmd\zeta \nonumber \\
+\frac{1}{2v}\int_{|\zeta|>\sqrt{v}}I_1(\beta,\zeta)(I_2(\beta,\zeta))^{v-1}
\rmd\zeta. \label{eq:5:16}
\end{eqnarray}
The function $M$ is bounded and satisfies 
\begin{equation*}
  |M(\beta,\zeta)|\leq\exp\left(\frac{2\sqrt{\beta\Lbar}}{\sqrt{\pi}}-\frac
{|\zeta|}{2}\right)
\end{equation*}
for $\zeta$ sufficiently large. Integrating \eref{eq:5:30} gives
\begin{align}
  \frac{1}{2v}\int_{-\sqrt{v}}^{\sqrt{v}} I_1(\beta,\zeta) 
(I_2(\beta,\zeta))^{v-1}& \rmd\zeta =
\frac{1}{2\sqrt{\pi}}\int_{-\sqrt{v}}^{\sqrt{v}}\frac{1}
{\sqrt{\rmi\beta\Lbar}}\exp\left(\frac{-\zeta^2}{4\rmi\beta\Lbar}\right)
M(\beta,\zeta)\rmd\zeta \nonumber \\
&+\Or\left[\int_{-\sqrt{v}}^{\sqrt{v}}\exp\left(-\frac{|\zeta|}{2}\right)
\left(\frac{1}{\sqrt{v}}+\frac{\zeta^2}{v}\right)\rmd\zeta\right]
\label{eq:5:35}
\end{align}
and the integral in the remainder term converges as we let $v\to\infty$.

To deal with the integral
\begin{equation*}
  \frac{1}{v}\int_{|\zeta|>\sqrt{v}}I_1(\beta,\zeta)(I_2(\beta,\zeta))^{v-1}
\rmd\zeta
\end{equation*}
we make use of the result of lemma \ref{lem:5:7}, giving
\begin{align}
\Bigg|\frac{1}{v}\int_{|\zeta|>\sqrt{v}}I_1(\beta,\zeta)
(I_2(\beta,\zeta&))^{v-1}\rmd\zeta\Bigg| \nonumber \\
&\leq \frac{1}{\pi}\int_{|\zeta|>\sqrt{v}} \left(
\frac{\beta\Lbar}{v^2}+\frac{\zeta^2}{\beta\Lbar}\right)^{-1}\rmd\zeta
+\Or(v^{-1}) \label{eq:5:41}\\ 
&\to0 \qquad\mbox{as $v\to\infty$.} \nonumber
\end{align}

Hence substituting \eref{eq:5:35} and \eref{eq:5:41} into \eref{eq:5:16}
and taking the limit $v\to\infty$ gives the required result.
\finire

\subsection{Properties of $M(\beta,\zeta)$}
We wish to rotate the variable $\zeta$. However, we need to check that 
the function $M$ does not blow up for large $|\zeta|$.

If $\zeta=R\rme^{\rmi\theta}$, then
\begin{equation}
  \erf\left(\frac{\zeta}{2\sqrt{\rmi\beta\Lbar}}\right)=\erf\left(
\frac{R}{2\sqrt{\beta\Lbar}}\rme^{\rmi(\theta-\pi/4)}\right)=1+\Or(R^{-1})
\label{eq:7:7}
\end{equation}
as $R\to\infty$, provided that $0<\theta<\pi/2$ \cite{abr:ste}. So,
\begin{eqnarray*}
  \left|\exp\left(-\zeta\erf\left(\frac{\zeta}{2\sqrt{\rmi\beta\Lbar}}\right)
\right)\right|&=&\exp\left(-R\;\Re\left[\rme^{\rmi\theta}\erf\left(\frac{\zeta}{2\sqrt{\rmi\beta
\Lbar}}\right)\right]\right)\\
&=&\rme^{-R\cos\theta}\Or(1)\\
&\to&0\qquad\mbox{as $R\to\infty$ provided $0<\theta<\pi/2$}
\end{eqnarray*}
and convergence is exponentially fast.
Similarly,
\begin{eqnarray*}
  \left|\exp\left(-\frac{2\sqrt{\rmi\beta\Lbar}}{\sqrt{\pi}}\exp
\left(-\frac{\zeta^2}{4\rmi\beta\Lbar}\right)\right)\right|
&=&\exp\left(\frac{-2\sqrt{\beta\Lbar}}{\sqrt{\pi}}\exp\left(-\frac{R^2}{4\beta
\Lbar}\sin2\theta\right)\right.\\
& &\left.\times\cos\left(\frac{R^2}{4\beta\Lbar}\cos2\theta+\frac{\pi}
{4}\right)\right)\\
&\to&1\qquad\mbox{as $R\to\infty$} 
\end{eqnarray*}
provided $0<\theta<\pi/2$.

Hence, if $0<\theta<\pi/4$ then
\begin{equation*}
 \lim_{R\to\pm\infty}  |RM(\beta,R\rme^{\rmi\theta})|=0
\end{equation*}
and we can make the change of variables $\zeta=\xi\sqrt{\rmi\beta\Lbar}$:
\begin{equation}
\int_{-\infty}^{\infty}
\frac{1}
{\sqrt{\rmi\beta\Lbar}}\exp\left(\frac{-\zeta^2}{4\rmi\beta\Lbar}\right)
M(\beta,\zeta)\rmd\zeta
=
\int_{-\infty}^{\infty}\rme^{-\xi^2/4}M(\beta,\xi\sqrt{\rmi\beta\Lbar})\rmd
\xi. \label{eq:5:10}
\end{equation}
We observe that
\begin{eqnarray*}
  M(\beta,\xi\sqrt{\rmi\beta\Lbar})&=&\exp\left(-\frac{2}{\sqrt{\pi}}
\sqrt{\rmi\beta\Lbar}\rme^{-\xi^2/4}-\xi\sqrt{\rmi\beta\Lbar}\erf
\left(\frac{\xi}{2}\right)\right)\\
&=&\exp\left(-\sqrt{\rmi\beta\Lbar}m(\xi)\right)
\end{eqnarray*}
where
\begin{equation*}
  m(\xi):=\frac{2}{\sqrt{\pi}}\rme^{-\xi^2/4}+\xi\erf(\xi/2).
\end{equation*}
\subsection{{\sl Proof of theorem \ref{thm:4}}}
$m$ satisfies the bound $m(\xi)\geq2/\sqrt{\pi}$, so
$M( \beta,\xi\sqrt{\rmi\beta\Lbar})$ is bounded for all $\beta$. 
By the Weierstrass $M$-test 
the integral in \eref{eq:5:10} is uniformly convergent and hence
\begin{equation*}
  \lim_{v\to\infty} E_v(\beta)
\end{equation*}
is a continuous function of $\beta$. We appeal, once again, to the 
continuity theorem for characteristic functions to deduce that the
limiting density $P(y)$ exists and is given by
\begin{eqnarray}
  P(y)&=&\frac{1}{4\pi^{3/2}}\int_{-\infty}^{\infty}\!\int_{-\infty}^{\infty}
\rme^{-\xi^2/4}M(\beta,\xi\sqrt{\rmi\beta\Lbar})\rme^{\rmi\beta y}\rmd
\xi\rmd\beta \nonumber \\
&=&\frac{1}{2\pi^{3/2}}\Re\int_{0}^{\infty}\!\int_{-\infty}^{\infty}
\rme^{-\xi^2/4}M(\beta,\xi\sqrt{\rmi\beta\Lbar})\rme^{\rmi\beta y}\rmd
\xi\rmd\beta.\label{eq:4:66}
\end{eqnarray}
The integrand is dominated by
\begin{equation*}
  \exp\left(-\xi^2/4-\sqrt{\pi\Lbar\beta}\right),
\end{equation*}
so Fubini's theorem allows us to switch the order of integration. We quote
the standard integral
\begin{equation*}
   \int_0^{\infty}\rme^{ax+b\sqrt{x}}\rmd{x}=-\frac{1}{a}-\frac{b}{2a}\sqrt
{\frac{\pi}{-a}}\exp\left(\frac{-b^2}{4a}\right)\erfc\left(\frac{-b}
{2\sqrt{-a}}\right)
\end{equation*}
valid for $\Re\: a<0$ and use this to perform the $\beta$ integral
in \eref{eq:4:66}. This leads to the result
\begin{equation*}
 P(y)=\frac{\sqrt{\Lbar}}{4\pi y^{3/2}}\Re\int_{-\infty}^{\infty}
\exp\left(-\frac{\xi^2}{4}-\frac{\Lbar m(\xi)^2}{4y}\right)m(\xi)
\erfc\left(\frac{\sqrt{\Lbar}m(\xi)}{2\rmi y}\right)\rmd\xi,
\end{equation*}
which reduces to the form given in the statement of the theorem upon
noticing that $\Re\{ \erfc(\rmi\theta)\}=1$ for all $\theta\in{\mathbb R}$.
\finire
\section{Value distribution of the eigenfunctions in the limit $v\to\infty$}
\label{sec:5}
\begin{figure}[h]
\begin{center}
\setlength{\unitlength}{7cm}
\begin{picture}(1,0.9)
\put(-0.25,0.1){\includegraphics[angle=0,width=9.0cm,height=5cm]{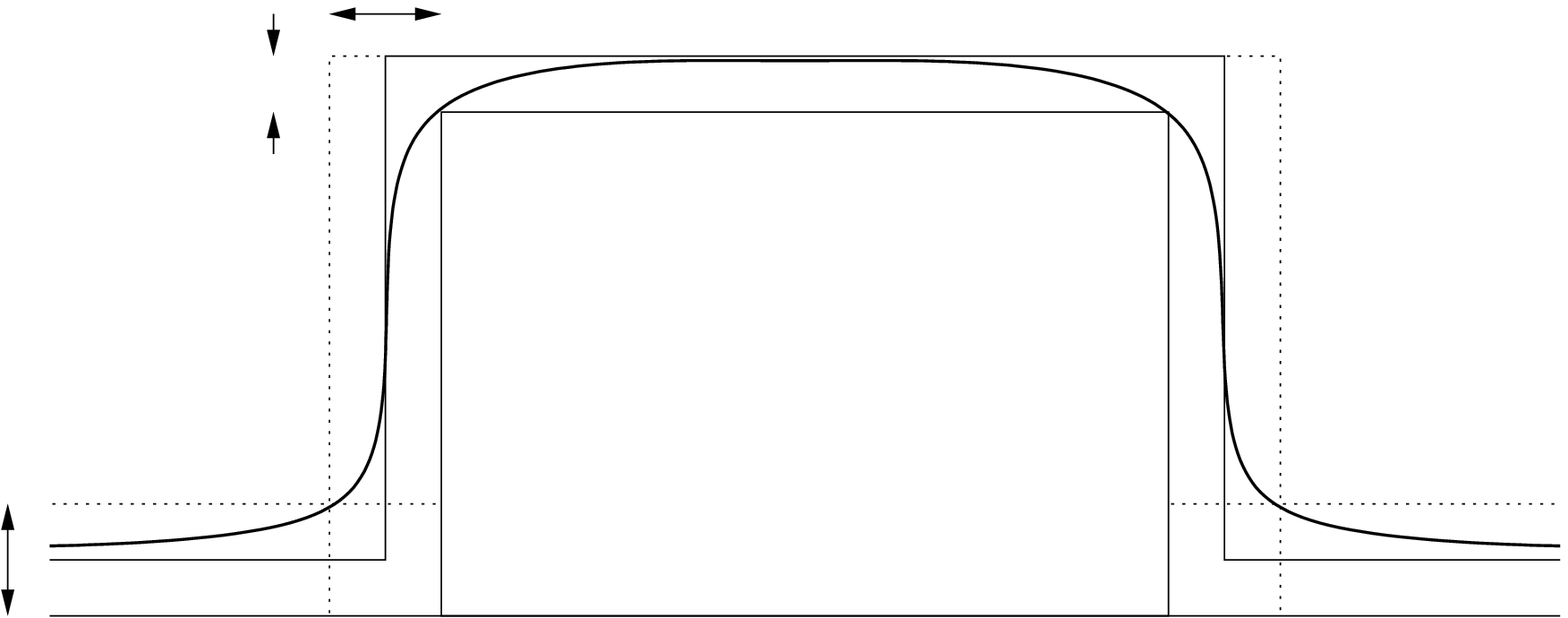}}
\put(-0.32,0.15){$2\sigma$}
\put(0.1,0.03){$a$}
\put(0.68,0.03){$b$}
\put(-0.09,0.7){$\sigma$}
\put(0.04,0.85){$2\sigma$}
\put(0.4,0.8){$\chi_1$}
\end{picture}
\caption{Approximating $\I_{[a,b]}$ from above}
\label{fig:6}
\end{center}
\end{figure}

To prove theorem \ref{thm:6} we use a standard approximation argument.
We introduce the smoothed $\delta$-function 
\begin{equation}
  \delta_{\epsilon}(x):=\frac{1}{2\pi}\int_{-\infty}^{\infty}
\rme^{-\epsilon|\beta|+\rmi\beta x}\rmd\beta=
\frac{\epsilon}{\pi(\epsilon^2+x^2)}.
\end{equation}
\begin{proposition}
\label{prop:5:100}
Let $\tilde{Q}_v(\eta)$ be related to $Q_v(\eta)$ by
\begin{equation*}
  \tilde{Q}_v(\eta):=\frac{1}{\eta^2}Q_v\left(\frac{1}{\eta}\right).
\end{equation*}
  For any fixed $\epsilon$,
  \begin{equation}
\lim_{v\to\infty}\int_0^{\infty}\delta_{\epsilon}(\eta-\eta')
\tilde{Q}_v(\eta')\rmd
\eta'=\int_0^{\infty} \delta_{\epsilon}(\eta-\eta')\tilde{Q}(\eta')\rmd\eta',
\label{eq:5:100}
  \end{equation}
uniformly for $\eta$ in compact intervals, where
\begin{equation*}
 \tilde{Q}(\eta)=\frac{1}{2\pi^{3/2}\eta}\Im\int_{-\infty}^{\infty}
\exp\left(-\frac{\xi^2}{4}
-\frac{\Lbar m(\xi)^2}{8\eta}\right)\erfc\left(\frac{\sqrt{\Lbar}m(\xi)}
{2\sqrt{2\eta}\rmi}\right)\rmd\xi.
\end{equation*}
\end{proposition}
\dimonstrazione
Let 
\begin{equation*}
  \tilde{Q}_{\epsilon,v}(\eta):=\int_{0}^{\infty}\delta_{\epsilon}(\eta-\eta')
\tilde{Q}_v(\eta')\rmd\eta'
=\lim_{N\to\infty}\frac{1}{N}\sum_{n=1}^{N}
\delta_{\epsilon}\left(\eta-\frac{1}{v^2A_i(n,\vec{L};v)}\right).
\end{equation*}
We use the identity 
\begin{equation*}
  \delta_{\epsilon}\left(\eta-\frac{A}{B}\right)=B\delta_{\epsilon|B|}(B\eta
-A)
\end{equation*}
with
\begin{equation*}
  A=\frac{\Lbar}{v^2}\sum_{j=1}^v \sec^2{kL_j}
\end{equation*}
and
\begin{equation*}
  B=2\sec^2kL_i.
\end{equation*}
Thus
\begin{eqnarray*}
  \delta_\epsilon\left(\eta-\frac{\sum_j 
\Lbar\sec^2 kL_j}{2v^2\sec^2kL_i}\right)
&=&\frac{2\sec^2kL_i}{\pi}\Re\int_{0}^{\infty}
\exp\left(\rmi\beta\left(2\eta-\frac{\Lbar}{v^2}+2\rmi\epsilon\right)
\sec^2{kL_i}\right)\\
& &\times
\prod_{j\neq i}\exp\left(-\frac{\rmi\beta\Lbar}{v^2}\sec^2 kL_j\right)
\rmd\beta.
\end{eqnarray*}
Applying identity \eref{eq:states}   to this function and following the 
method in section \ref{sec:5} leads to
\begin{align*}
  \frac{2\Lbar}{\pi^2\bar{d}v}\Re\int_{0}^{\infty}\int_{-\infty}^{\infty}
I_4(\beta,\zeta)&((I_2(\beta,\zeta))^{v-1}\\
&+(v-1)I_3(\beta,\zeta)I_1(\beta,\zeta)
(I_2(\beta,\zeta))^{v-2}\rmd\zeta\rmd\beta
=:\tilde{P}_{\epsilon}(\eta)
\end{align*}
where the new integrals are
\begin{equation*}
   I_3(\beta,\zeta):=\frac{1}{\pi}\int_{0}^{\pi}\sec^2x\exp\left(\rmi\beta
\left(2\eta-\frac{\Lbar}{v^2}+2\rmi\epsilon\right)\sec^2x+
\frac{\rmi\zeta}{v}\tan x
\right)\rmd x
\end{equation*}
and
\begin{equation*}
  I_4(\beta,\zeta):=\frac{1}{\pi}\int_{0}^{\pi}\sec^4x\exp\left(\rmi\beta
\left(2\eta-\frac{\Lbar}{v^2}+2\rmi\epsilon\right)
\sec^2x+\frac{\rmi\zeta}{v}\tan x
\right)\rmd x.
\end{equation*}
$I_3$ and $I_4$ converge uniformly in $\zeta$ and $\eta$.
Integral $I_3$ closely resembles integral $I_1$.
\begin{equation}
  I_3(\beta,\zeta)=\frac{\rme^{2\rmi\beta\eta-2\epsilon\beta}}{\sqrt{\beta}
\sqrt{2\pi\epsilon-2\pi\rmi\eta}}+\Or_{\epsilon}(v^{-1})\qquad\mbox{as $v\to\infty$.}
\end{equation}
By making the substitution $z=\tan x$, $I_4$ reduces to
\begin{eqnarray*}
  I_4(\beta,\zeta)&=&\frac{1}{\pi}\int_{-\infty}^{\infty} (1+z^2)\exp\left(
\rmi\beta\left(2\eta-\frac{\Lbar}{v^2}+2\rmi\epsilon\right)
(1+z^2)+\frac{\rmi\zeta z}{v}
\right)\rmd z\\
&=&\Or_{\epsilon}(1)\qquad\mbox{as $v\to\infty$.}
\end{eqnarray*}
This estimate ensures that $\tilde{Q}_{\epsilon,v}(\eta)$ is dominated by
the second term. Since $I_3$ is also bounded as $v\to\infty$, the
analysis of proposition \ref{prop:5:9} holds and
\begin{align*}
&\lim_{v\to\infty}\tilde{Q}_{\epsilon,v}(\eta)\\
&=\frac{2}{\pi^2}\Re
\frac{1}{\sqrt{2\epsilon-2\rmi\eta}}
\int_0^{\infty}\!
\int_{-\infty}^{\infty}\frac{1}{\beta\sqrt{\rmi\Lbar}}\exp\left
(2\rmi\beta\eta-\frac{\zeta^2}
{4\rmi\beta\Lbar}-2\epsilon\beta\right)
 M(\beta,\zeta)\rmd\zeta\rmd\beta\\
&=\frac{2}{\pi^2}\Re\left[\frac{1}{\sqrt{2\epsilon-2\rmi\eta}}
\int_{-\infty}^{\infty}\int_0^{\infty}
\exp\left(-\frac{\xi^2}{4}-\sqrt{\rmi\beta\Lbar}m(\xi)
+2\rmi\beta\eta-2\epsilon\beta
\right)\frac{\rmd\beta}{\sqrt{\beta}}\rmd\xi\right],
\end{align*}
where we have made, once again, the substitution 
$\zeta=\xi\sqrt{\rmi\beta\Lbar}$
in the final line. Putting $\theta^2=\beta$ reduces the $\beta$ integral to
\begin{align*}
  \int_0^{\infty}\rme^{-\sqrt{\rmi\Lbar}\theta m(\xi)-(2\epsilon-2\rmi\eta)
\theta^2}
\rmd\theta\mspace{-100mu}&\\ 
&= \frac{\sqrt{\pi}}{2\sqrt{2\epsilon-2\rmi\eta}}
\exp\left(\frac{-\Lbar m(\xi)^2}
{8(\eta+\rmi\epsilon)}\right)
\erfc\left(\frac{\sqrt{\Lbar}m(\xi)}{2\rmi
\sqrt{2\eta+2\rmi\epsilon}}\right),
\end{align*}
so
\begin{align*}
\lim_{v\to\infty}&\tilde{Q}_{\epsilon,v}(\eta)\\
&=\frac{1}{2\pi^{3/2}}\Im\left[
\frac{1}{\eta+\rmi\epsilon}
\int_{-\infty}^{\infty}\exp\left(-\frac{\xi^2}{4}-\frac{\Lbar m(\xi)^2}
{8(\eta+\rmi\epsilon)}\right)\erfc\left(\frac{\sqrt{\Lbar}m(\xi)}{2\rmi
\sqrt{2\eta+2\rmi\epsilon}}\right)\rmd\xi\right]\\
&=\int_{0}^{\infty}\delta_{\epsilon}(\eta-\eta')\tilde{Q}(\eta')\rmd\eta'
\end{align*}
\finire
We note that $\tilde{Q}(\eta)$ is a continuous probability density on 
$(0,\infty)$.
\dimon{Proof of theorem \ref{thm:6}.}
Let $0<a<b$ be fixed.
Then let
\begin{equation*}
  \I^{\epsilon}_{[a,b]}(\eta):=\int_{a}^{b} 
\delta_{\epsilon}
(\eta-y)\rmd y.
\end{equation*}
$\I^{\epsilon}_{[a,b]}(\eta)$ converges pointwise as 
$\epsilon\to 0$ to
the function $\I_{[a,b]}(\eta)$ everywhere
apart from at the end-points $a$ and $b$. Given $\sigma>0$, consider the
function
\begin{equation*}
  \chi_1(\eta):=\I^{\epsilon}_{[a-\sigma,b+\sigma]}(\eta)+\sigma.
\end{equation*}
There exists $\epsilon>0$ such that:-
\begin{center}
\begin{itemize}
  \item $0\leq\chi_1(\eta)\leq 2\sigma$ for 
$\eta\leq a-2\sigma$ and $\eta\geq b+2\sigma$,
  \item $1\leq\chi_1(\eta)\leq 1+\sigma$ for 
$a\leq\eta\leq b$,
 \item $0\leq\chi_1(\eta)\leq 1+\sigma$ for 
$a-2\sigma\leq\eta\leq a$ and $b\leq\eta\leq b+2\sigma$.
\end{itemize}
\end{center}
This construction is illustrated in figure \ref{fig:6}

Similarly, the function
\begin{equation*}
\chi_2(\eta)=\I^{\epsilon}_{[a+\sigma,b-\sigma]}(\eta)-\sigma
\end{equation*}
satisfies for $\epsilon$ sufficiently small:-
\begin{itemize}
  \item $-\sigma\leq\chi_2(\eta)\leq 0$ for $\eta\leq a$
and $\eta\geq b$,
  \item $1-2\sigma\leq\chi_2(\eta)\leq 1$ for 
$a+2\sigma\leq\eta\leq b-2\sigma$,
  \item $-\sigma\leq\chi_2(\eta)\leq 1$ for all 
$a\leq\eta\leq a+2\sigma$ and $b-2\sigma\leq\eta\leq b$.
\end{itemize}
So that for all $\eta\in[0,\infty)$
  \begin{equation}
  \chi_2(\eta)<\I_{[a,b]}(\eta)<\chi_1(\eta).
\label{eq:5:69}
  \end{equation}
Also,
\begin{eqnarray*}
  \int_0^{\infty}\left[\chi_1(\eta)-
\chi_2(\eta)\right]\tilde{Q}(\eta)\rmd\eta\leq
3\sigma\int_{0}^{\infty}\tilde{Q}(\eta)\rmd\eta+(1+2\sigma)
\int_{a-2\sigma}^{a
+2\sigma}\tilde{Q}(\eta)\rmd\eta\\
+(1+2\sigma)\int_{b-2\sigma}^{b+2\sigma}\tilde{Q}(\eta)\rmd\eta,
\end{eqnarray*}
which can be made arbitrarily small because $\tilde{Q}$ is a continuous 
probability density. It follows from proposition \ref{prop:5:100} that
\begin{equation*}
\lim_{v\to\infty}\int_0^{\infty}\chi_1(\eta)\tilde{Q}_v(\eta)
\rmd\eta=\int_0^{\infty}\chi_1(\eta)\tilde{Q}(\eta)\rmd\eta
\end{equation*}
and similarly for $\chi_2$. Hence, we can use the argument
of lemma \ref{lem:2} {\em mutatis mutandis}, to deduce that 
\begin{equation}
 \lim_{v\to\infty}\int_a^b\tilde{Q}_v(\eta)\rmd\eta=\int_{a}^{b}\tilde{Q}(\eta)
\rmd\eta.
\end{equation}
Making the substitution $\eta\mapsto 1/\eta$ then completes the proof
of convergence.

Expanding the error function in \eref{eq:Q:def} as
\begin{equation*}
  \erfc\left(\frac{\sqrt{\Lbar\eta}m(\xi)}{2\rmi\sqrt{2}}\right)=
\frac{1}{\sqrt{\pi}}\exp\left(\frac{\Lbar\eta m(\xi)^2}{8}\right)
\left(\frac{2\rmi\sqrt{2}}{\sqrt{\Lbar\eta}m(\xi)}+\Or(\eta^{-3/2})\right),
\end{equation*}
where the implied constant does not depend on $\xi$, yields
\begin{equation*}
  Q(\eta)=\frac{b}{\eta^{3/2}}+\Or(\eta^{-5/2})\qquad
\mbox{as $\eta\to\infty$,}
\end{equation*}
where the constant $b$ is
\begin{equation*}
b=\frac{\sqrt{2}}{\sqrt{\Lbar}\pi^2}\int_{-\infty}^{\infty}
\frac{\rme^{-\xi^2/4}}{m(\xi)}\rmd\xi
\approx\frac{0.348}{\sqrt{\Lbar}}.
\end{equation*}
\finire
The algebraic decay of $Q(\eta)$ is in contrast to the 
exponential decay of the $\chi_1^2$ density. 

\section{Numerical Results}
\label{sec:6}
The results presented above show close agreement with numerical computations.
We present these computations now by way of illustration.

In all the figures in this section, the choice of $\Lbar=2$ has been made.

\begin{figure}[h]
\begin{center}
\includegraphics[angle=0,width=8.0cm,height=6cm]{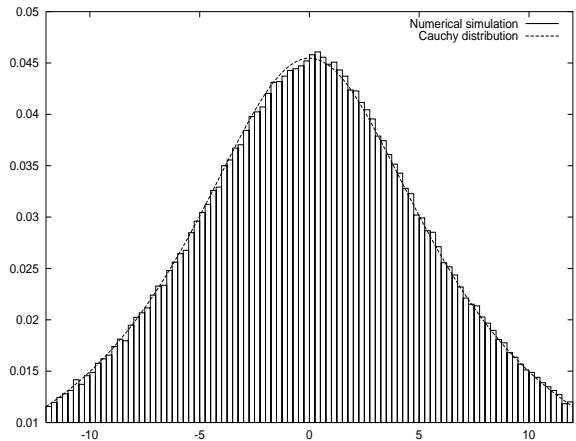}
\caption{The value distribution for the spectral determinant}
\label{fig:9:3}
\end{center}
\end{figure}

Figure \ref{fig:9:3} shows a comparison between a numerical evaluation of 
values taken by the spectral determinant and the Cauchy distribution.
The numerical evaluation was based on a star graph with 7 randomly chosen
bond lengths, and 100,000 samples of $k$.

\begin{figure}[h]
\begin{center}
\includegraphics[angle=0,width=8.0cm,height=6cm]{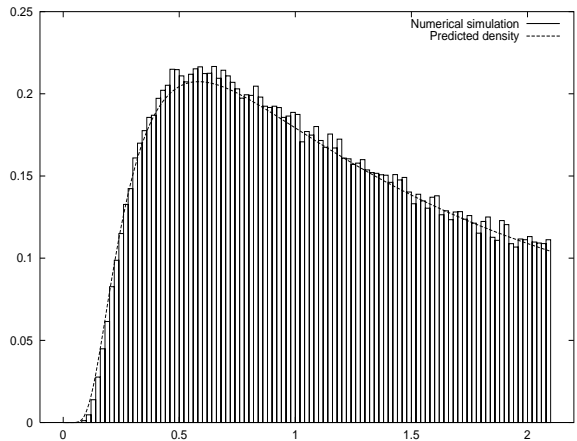}
\caption{The value distribution of $Z'(k)$}
\label{fig:9:4}
\end{center}
\end{figure}

Figure \ref{fig:9:4} shows a comparison between the distribution of values
taken by the derivative of the spectral determinant at its zeros, and the
corresponding numerical evaluation. Plotted is numerical data for a 70-bond 
star
graph, together with the $v\to\infty$ limiting density given in theorem
\ref{thm:4}. Once again we see good agreement.

\begin{figure}[h]
\begin{center}
\includegraphics[angle=0,width=8.0cm,height=6cm]{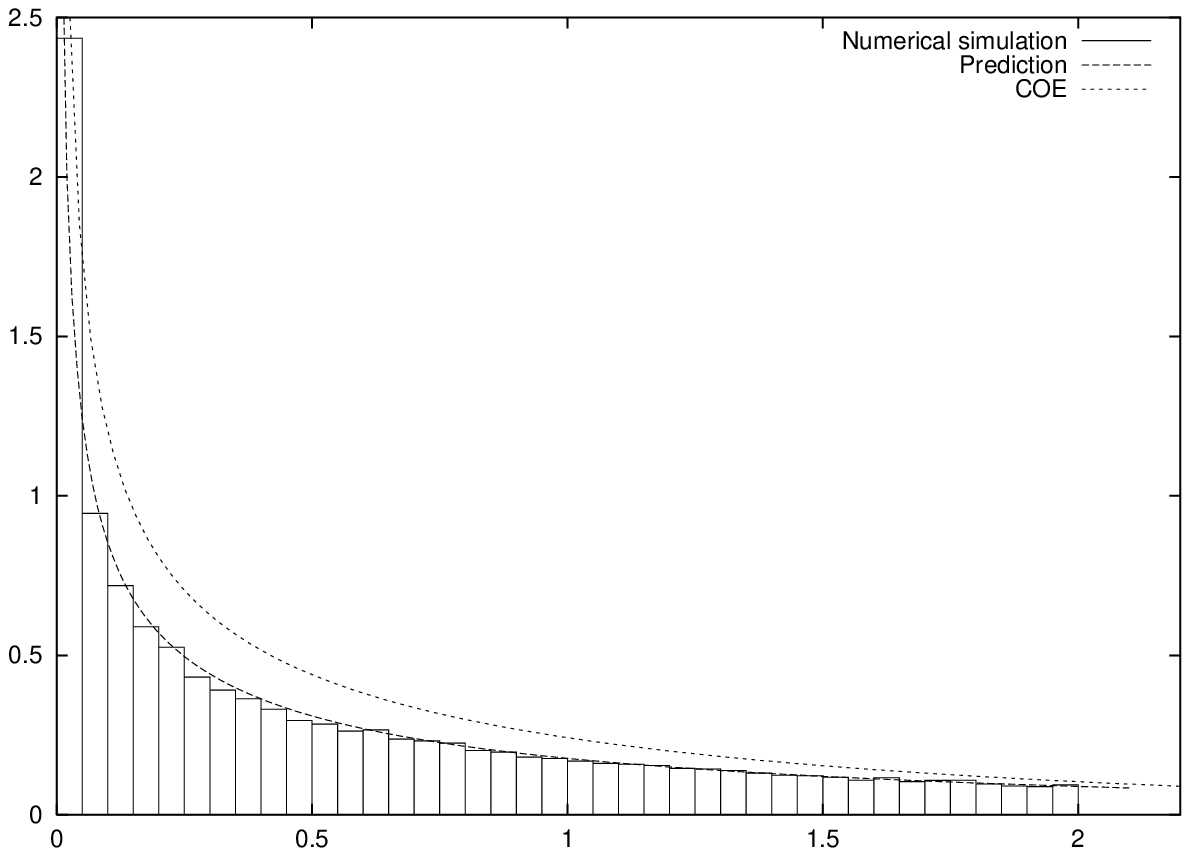}
\caption{The value distribution of $A_i(n,\vec{L};v)$}
\label{fig:9:5}
\end{center}
\end{figure}

In figure \ref{fig:9:5} we compare a numerical evaluation of the density of
values taken by the maximum norm of eigenvectors of a 50-bond graph
to the $v\to\infty$ limiting density given in theorem \ref{thm:6}.
Also plotted for comparison is the density of the $\chi^2_1$ distribution 
associated with the COE of random matrices.

\section{Connections with the \v{S}eba Billiard}
\label{sec:7}
The correspondence between the spectral statistics of quantum star graphs
and those of \v{S}eba billiards with periodic boundary conditions has 
already been noted \cite{berk:2}. This
is due to the fact that the spectral determinant for the star graphs 
\eref{spec:det} may be re-written in a form similar to the spectral determinant
of a \v{S}eba billiard:
\begin{equation*}
Z_{\rm Seba}(E)=\sum_{k=1}^{\infty}\frac{1}{E_k^{(0)}-E}
\end{equation*}
where the $E_k^{(0)}$ are the energy levels of the unperturbed system. Both
spectral determinants have infinitely many poles of first order, 
which separate the 
energy levels of the perturbed system. We therefore expect the 
value distribution of the spectral determinant of a \v{S}eba billiard
to be Cauchy to be consistent with
theorems \ref{thm:1} and \ref{thm:2}.

\begin{figure}[h]
\begin{center}
\includegraphics[angle=0,width=8.0cm,height=6cm]{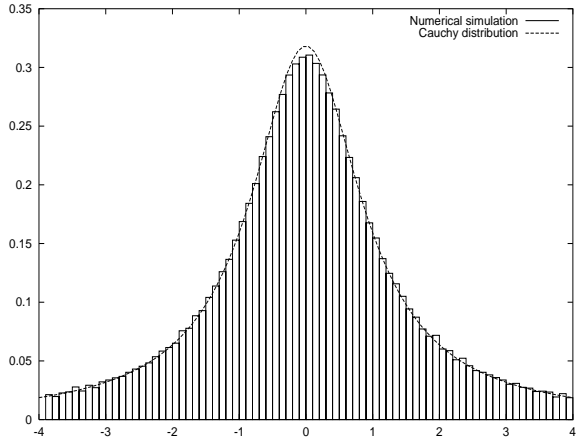}
\caption{The value distribution for the spectral determinant of a 
\v{S}eba billiard}
\label{fig:10:1}
\end{center}
\end{figure}

This conjecture is supported by figure \ref{fig:10:1} which is a plot of the
density of values given taken by the function
\begin{equation}
  \pi\langle d\rangle\sum_{k=1}^{K}\frac{1}{E_k^{(0)}-E}
\label{eq:conj:1}
\end{equation}
for $K=3000$ unperturbed levels of a rectangular quantum billiard with
Neumann boundary conditions, with $E$ 
distributed uniformly between
$E_{1000}^{(0)}$ and $E_{2000}^{(0)}$. The constant $\langle d\rangle$ is the
mean density of levels of the system and
it takes the place of the constant $v^{-1}$ in 
theorems \ref{thm:1} and \ref{thm:2}.
The fit to a Cauchy density is convincing.

If we treat the unperturbed levels in \eref{eq:conj:1} as  independent
identically distributed random variables with a uniform density then the random
variables
\begin{equation*}
  \frac{1}{E_k^{(0)}-E}
\end{equation*}
have a distribution that falls into the domain of attraction of the stable
Cauchy density. That the limiting density is Cauchy is then a classical result
of probability theory \cite{fel:ipt}.

We now present an argument
which suggests that the normalisation constant associated with the wave 
functions of the 
\v{S}eba billiard also shares significant features with the normalisation 
constant of the star graphs \eref{norm:const}.

The wave functions of a general \v{S}eba billiard \cite{seba:2} can be
written in the form
\begin{equation}
\psi_n(\vec{x})=A_n\sum_{k=1}^{\infty} \frac{\psi_k^{(0)}(\vec{x}_0)
\psi_k^{(0)}(\vec{x})}{E_k^{(0)}-E_n}
\end{equation}
where $E_n$ is the $n^{\rm th}$ energy level and $E_k^{(0)}$ and 
$\psi_k^{(0)}$ are, respectively, the energy levels
and wave functions of the original integrable system of which the \v{S}eba
problem is a perturbation.
The Berry-Tabor Conjecture \cite{berry:1} asserts that the unperturbed
levels $E_k^{(0)}$ are distributed like Poisson variables; that is independent
and random.  We fix the usual normalisation
\begin{equation*}
\int|\psi_n(\vec{x})|^2 \rmd\vec{x}=1,
\end{equation*}
which leads to a value for the constant $A_n$
\begin{equation}
A_n^2=\left(\sum_{k=1}^{\infty} \frac{|\psi_k^{(0)}(\vec{x}_0)|^2}
{(E_k^{(0)}-E_n)^2}\right)^{-1}.
\label{seba:norm}
\end{equation}
In the case that the unperturbed system is a rectangular quantum billiard with
Neumann boundary conditions and sides of length $\alpha^{1/4}$ and
$\alpha^{-1/4}$
the wavefunctions are
\begin{equation}
\psi_{n,m}^{(0)}(x,y)=2\cos\left(\frac{n\pi x}{\alpha^{1/4}}\right)
\cos(m\pi y\alpha^{1/4}),\qquad\mbox{$n,m=0,1,2,\ldots$.}
\end{equation}
If we position the scatterer at the origin, then 
$|\psi_{n,m}^{(0)}(\vec{x}_0)|=2$. 
This billiard problem is equivalent to the
billiard with periodic boundary conditions desymmetrised to remove degeneracies
in the spectrum.
Provided that the constant
$\alpha$ satisfies certain diophantine conditions (see 
\cite{marklof:1,eskin:1} for
details) then $A_n^2$ in \eref{seba:norm} is the reciprocal of a sum
of functions with poles of second order distributed independently. 
These poles play the r\^ole of the singularities of the functions $\sec kL_j$
which appear in the normalisation of the quantum graphs. Such poles 
determine the rate of decay of the tails of the relevant probability
distributions, and this implies that the analysis performed
in the present work also holds for this billiard
problem. In particular, we conjecture that the distribution
of the square of the $i^{\rm th}$ coefficient of the 
eigenfunctions in the basis 
$|\psi_k^{(0)}\rangle$ is the same as the limiting distribution of 
$A_i(n,\vec{L};v)$.

\begin{figure}[h]
\begin{center}
\includegraphics[angle=0,width=8.0cm,height=6cm]{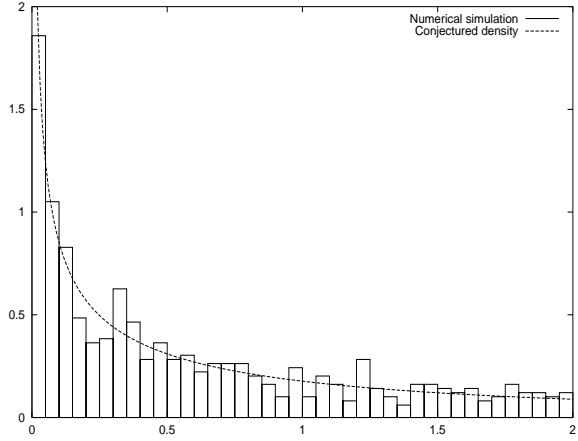}
\caption{The value distribution for the eigenfunctions of a \v{S}eba billiard}
\label{fig:10:2}
\end{center}
\end{figure}

We present in figure \ref{fig:10:2} the distribution of values taken by
\begin{equation*}
  c\frac{(E_i^{(0)}-E_n)^{-2}}{\sum_{k=1}^{K} (E_k^{(0)}-E_n)^{-2}}
\end{equation*}
where $n$ is now a random variable uniformly distributed on
 $\{1000,\ldots,2000\}$ and
we take $K=3000$, $i=1500$ and $\alpha=(\sqrt{5}-1)/2$. 
The constant $c$, which in general may be expected to depend on
$K$ and the distribution of $n$, is required to ensure that the 
sum of terms in the denominator is normalised and to compensate for the fact
that the functions are not periodic. In order to compare with the corresponding
results for star graphs we require that the tail of the 
distribution of $c(E_i^{(0)}-E_n)^{-2}$ is asymptotic to the tail
of the distribution of $(2/\Lbar)\sec^2k_nL_i$.
Assuming $n$ to be distributed between $n_{\rm max}$ and $n_{\rm min}$,
a heuristic examination of
these densities leads to the association
\begin{equation*}
  c=\frac{2}{\Lbar}{(E_{\rm max}-E_{\rm min})^2}{\langle d\rangle^2}
\end{equation*}
where $E_{\rm max}$ and $E_{\rm min}$ are respectively the energy levels
corresponding to $n=n_{\rm max}$ and $n=n_{\rm min}$.
For the data in figure \ref{fig:10:2}, we get $c\approx9.75\times10^5$.

\section*{Acknowledgements}
JM is supported by an EPSRC Advanced
Research Fellowship, the Nuffield Foundation (Grant NAL/00351/G), and
the Royal Society (Grant 22355).
BW is supported by an EPSRC studentship (Award Number 0080052X).
Additionally, we are grateful for the financial support of the European 
Commission 
under the Research Training Network (Mathematical Aspects of Quantum Chaos) 
HPRN-CT-2000-00103 of the IHP Programme.

\appendix
\section*{Appendix}

We here show that the distribution of the maximum amplitude $A_i(n,L;v)$
completely determines the value distribution
of the eigenfunctions on the $i^{\rm th}$ bond, which is described by
\begin{equation}\label{one}
\frac{1}{N L_i} \sum_{n=1}^N \int_0^{L_i} f\big( \psi_i^{(n)}(x)\big) \, \rmd x
\end{equation}
where $f$ is an arbitrary bounded continuous function.
Let us in fact consider the more general joint distribution,
\begin{equation*}
\frac{1}{N L_i} \sum_{n=1}^N \int_0^{L_i} 
F\big( \cos k_n(x-L_i),v^2 A_i(n,L;v)\big) \, \rmd x
\end{equation*}
where $F$ is a bounded continuous function in two variables.
We obtain the expression (\ref{one}) for the choice
$F(t,\eta)=f(t \sqrt{\eta})$ 
provided $f$ is even (which, as will become clear below, 
we may assume w.l.o.g.).

We begin with the special case when $F$ factorizes, i.e., 
$F(t,\eta)=f_1(t)\,f_2(\eta)$ where $f_1,f_2$ are arbitrary
bounded continuous functions.
Then 
\begin{align}\nonumber 
\frac{1}{L_i} \int_0^{L_i} 
f_1\big( \cos k_n(x-L_i)\big) \, \rmd x
&= \int_0^1 f_1\big(\cos(2\pi x)\big)\, \rmd x + \Or(k_n^{-1}) \\
&= \frac1\pi \int_{-1}^1 f_1(t) \frac{\rmd t}{\sqrt{1-t^2}} + \Or(k_n^{-1}),
\label{uno}
\end{align}
and, by theorem \ref{thm:5},
\begin{equation}\label{due}
\frac{1}{N} \sum_{n=1}^N f_2\big(v^2 A_i(n,L;v)\big) 
\to \int_0^\infty f_2(\eta) Q_v(\eta)\, \rmd\eta 
\end{equation}
as $N\to\infty$.
Since the mean density of the eigenvalues $k_n$ is constant, we have
$ \sum_{n\leq N} \Or(k_n^{-1}) = \Or(\log N)$ and thus
from (\ref{uno}) and (\ref{due})
\begin{multline} \label{final}
\lim_{N\to\infty}\frac{1}{N L_i} \sum_{n=1}^N \int_0^{L_i} 
F\big( \cos k_n(x-L_i),v^2 A_i(n,L;v)\big) \, \rmd x \\
=
\frac1\pi \int_{-1}^1  \int_0^\infty F(t,\eta) \,
Q_v(\eta)\, \frac{\rmd\eta\, \rmd t}{\sqrt{1-t^2}} .
\end{multline}
This holds for functions $F=f_1\, f_2$ and, by linearity, also
for finite linear combinations of such functions. 
Given any $\epsilon>0$ we can approximate any
bounded continuous $F$ from above and below by such finite
linear combinations $F_+$ and $F_-$, respectively, such that
$$
\frac1\pi \int_{-1}^1  \int_0^\infty \big[F_+(t,\eta)-F_-(t,\eta)\big] \,
Q_v(\eta)\, \frac{\rmd\eta\, \rmd t }{\sqrt{1-t^2}} < \epsilon .
$$
Since $\epsilon$ can be arbitrarily small, (\ref{final}) holds
in fact for any bounded continuous $F$. 

We can therefore choose $F(t,\eta)=f(t \sqrt{\eta})$ 
as a test function, and 
we find
\begin{equation*}
\lim_{N\to\infty}
\frac{1}{N L_i} \sum_{n=1}^N \int_0^{L_i} f\big( \psi_i^{(n)}(x)\big) \, \rmd x
=\int_{-\infty}^\infty f(r) \, R_v(r)\, \rmd r
\end{equation*}
with the limiting distribution
\begin{equation*}
R_v(r)=\frac{1}{\pi} \int_{r^2}^{\infty} 
Q_v(s) \frac{\rmd s}{\sqrt{s-r^2}}.
\end{equation*}

The limit $v\to\infty$ can be handled in an analogous way and leads to
the same formulas for the limit $R(r)$ of $R_v(r)$ with $Q_v(s)$
replaced by $Q(s)$ in the above. It follows from the asymptotic expansion
\eref{eq:Q:asympt} that $R(r)$ also decays with an algebraic tail.

\end{document}